\documentclass[apj]{emulateapj}
\usepackage{natbib}
\usepackage{color}

\submitted{DRAFT VERSION JUL. 15, 2016}
\shortauthors{H. Lee, A. Lazarian \& J. Cho}
\shorttitle{SYNCHROTRON POLARIZATION}

\begin{document}

\title{Polarimetric studies of magnetic turbulence with interferometer}

\author{Hyeseung Lee\altaffilmark{1}, A. Lazarian\altaffilmark{2}, and Jungyeon Cho\altaffilmark{1,2}}
\affil{1. Department of Astronomy and Space Science, Chungnam National University,Deajeon, Republic of Korea}
\affil{2. Department of Astronomy, University of Wisconsin, Madison, USA}

\begin{abstract}
We study statistical properties of synchrotron polarization emitted from media with magnetohydrodynamic (MHD) turbulence.
We use both synthetic and MHD turbulence simulation data for our studies.
We obtain the spatial spectrum and its derivative with respect to wavelength of synchrotron polarization arising from both synchrotron radiation and Faraday rotation fluctuations.
In particular, we investigate how the spectrum changes with frequency.
We find that our simulations agree with the theoretical predication in Lazarian \& Pogosyan (2016).
We conclude that the spectrum of synchrotron polarization and it derivative can be very informative tools to get detailed information about the statistical properties of MHD turbulence from radio observations of diffuse synchrotron polarization.
Especially, they are useful to recover the statistics of turbulent magnetic field as well as turbulent density of electrons. 
We also simulate interferometric observations that incorporate the effects of noise and finite telescope beam size, and demonstrate how we recover statistics of underlying MHD turbulence.
\end{abstract}


\section{INTRODUCTION}
\label{sect:intro}

Turbulence with embedded magnetic field is almost everywhere in the universe on a wide variety of scales, such as interstellar medium (Elmegreen \& Falgarone 1996; Elmegreen \& Scalo 2004) and intracluster medium (Ensslin \& Vogt 2006; Lazarian 2006). Substantial theoretical progress in relating MHD turbulence with  astrophysical processes has been made. The areas  include star formation (Larson 1981; McKee \& Tan 2002; Elmegreen 2002; Mac Low \& Klessen 2004; Ballesteros-Paredes et al. 2007; McKee \& Ostriker 2007), accretion disks (Balbus \& Hawley 2002), the solar wind (Hartman \& McGregor 1980; Podesta 2006; Wicks et al. 2012), magnetic reconnection (Lazarian \& Vishniac 1999; Lazarian et al. 2015) and cosmic rays (Schlickeiser 2002).

Turbulence is a chaotic phenomenon. However, it allows for a very simple statistical description. The landmark achievements include the famous Kolmogorov statistical theory (Kolmogorov 1941) as well as Goldreich \& Sridhar MHD turbulence theory (Goldreich \& Sridhar 1995). The latter is the theory relevant to most magnetized astrophysical fluids, including magnetic fields responsible for the most of the Galactic and extragalactic synchrotron emission\footnote{We note parenthetically that the first attempts to formulate the MHD turbulence theory can be traced back to the classical works of Iroshnikov (1964) and Kraichnan (1968). Later advances include Montgomery \& Turner (1981), Shebalin et al. (1983), and Higdon (1984). For the further advancements of the theory and its testing one can refer to a number of papers that include Lazarian \& Vishniac (1999), Cho \& Vishniac (2000), Maron \& Goldreich (2001), Cho et al. (2002). The extension of MHD turbulence theory for compressible turbulence can be found in Lithwick \& Goldreich (2001), Cho \& Lazarian (2002a,b, 2003), and Kowal \& Lazarian (2007). Recent reviews on the subject include Brandenburg \& Lazarian (2013) and Beresnyak \& Lazarian (2015).}.

As no in situ measurements of turbulence is possible beyond the very limited volume of the interplanetary medium and the solar wind, it is challenging to obtain turbulence statistics from observations. This area has been a focus of intensive theoretical and observational research for a number of decades with a significant progress achieved recently  (Muhch 1958; Munch \& Wheelon 1958; Burkhart et al 2012; Brunt \& Heyer 2013; Chepurnov et al. 2010, 2015). A lot of studies have focused on the velocity field statistics (see Lazarian 2009 for a review and references therein). However, for studies of MHD turbulence it is even more important to know the statistics of the complementary measure, namely, magnetic field. In this paper, we discuss how to obtain the statistical properties of magnetic field from observations of synchrotron polarization.

The simplest possible model of turbulence is Kolmogorov phenomenology, which provides a simple scaling relation for both incompressible and mildly compressible hydrodynamic turbulence cascade (Kolmogorov 1941, 1962). Suppose that energy is injected at a scale L, called the energy injection scale or the outer scale of turbulence. Then, the injected energy cascades to smaller scales with negligible energy losses and reaches the dissipation scale. The range between the energy injection scale and the dissipation scale is called inertial range and the Kolmogorov model predicts a $P_{3D}(\textbf{k})=P_{3D}(k) \propto k^{-11/3}$ three-dimensional (3D) power spectrum in the inertial range. Here \textbf{k} is the wave-vector in 3D space, i.e., $\textbf{k}=(k_x,k_y,k_z)$, and $k=\sqrt{k_x^2+k_y^2+k_z^2}$. The spectra of many astrophysical quantities typically exhibit 3D power spectra with power law indices close to $-11/3$ (Armstrong et al. 1995; Leamon et al 1998; Chepurnov \& Lazarian 2009; Lazarian et al. 2002), corresponding to Kolmogorov spectrum. If a quantity $s$ has a 3D power spectrum $k^{m}$ and we observe the quantity $S$ that is integrated along the line of sight (LOS), i.e.,~$S=\int s dl$, then the quantity $S$ has a 2D power spectrum of $P_{2D}(\textbf{K})=P_{2D}(K) \propto K^{m}$ (i.e.,~$|\tilde{S}_{\textbf{K}}|^2 \propto K^{m}$), where $\tilde{S}_{\textbf{K}}$ is the 2D Fourier transform of $S$, and \textbf{K} is the wave-vector in 2D space. If the LOS is along the z-direction, then $\textbf{K}=(k_x,k_y)$ and $K=\left|\textbf{K}\right|=\sqrt{k_x^2+k_y^2}$. In this case, the 1D spectrum for the 2D data $E_{2D}(K)$ (see Appendix B for definition) is proportional to $K^{m+1}$.

When the eddy motions perturb magnetic field lines and produce magnetic fluctuations, the perturbations leave imprints of turbulence statistics on magnetic field. Therefore, we can study turbulence by observing statistics of magnetic field. One of the easiest way to estimate the magnetic field direction and its strength in the plane perpendicular to the LOS (i.e., the plane of the sky) is via polarization resulting from synchrotron radiation. Relativistic electrons interact with magnetic field and emit synchrotron radiation, which is sensitive to both the strength and direction of magnetic field. The total intensity of synchrotron emission depends on the number density of electrons and the strength of the magnetic field perpendicular to the LOS: $I \propto \int N_{0}B_{\perp}^{(p +1)/2}\omega ^{-(p-1)/2} dz$, where the LOS is along the $z$ direction, $\omega$ is gyro-frequency, and p is the power-law index of electron energy distribution. In addition, since the synchrotron radiation is linearly polarized and the direction of polarization is perpendicular to the plane-of-the-sky magnetic field, we can infer the statistics of the plane-of-the-sky magnetic field from observations of synchrotron polarization. However,  Faraday rotation of  the direction of polarization makes the interpretation of synchrotron polarization more complicated. Faraday rotation can be obtained from measurements of the polarization angles $\chi$ at several wavelengths and quantified by the rotation measure (RM) defined by integration of electron density times the strength of the LOS magnetic field along the LOS. Since polarized synchrotron emission and Faraday rotation are related with magnetic field perpendicular to and parallel to the LOS, respectively, they can yield the information about the corresponding components of magnetic field in the region. In this paper, we concentrate on the statistical properties of synchrotron polarization.

The spectrum of polarized synchrotron emission is of great importance for understanding of many astrophysical phenomena (see Beck 2015).
The observed spectra of synchrotron emission and polarization reveal a range of power-law spectra (de Oliveira-Costa et al. 2003). It is clear that the spectrum of polarized synchrotron emission reflects that of magnetic fluctuations. This was usually demonstrated for the particular index of galactic cosmic rays, namely $\gamma~(\equiv (p+1)/2)=2$, which provides the dependence of the synchrotron signal proportional to magnetic field squared. More recently, the statistics of fluctuations for arbitrary $\gamma$ was obtained in Lazarian \& Pogosyan (2012; hereinafter LP12), where it was found how the change of $\gamma$ affects the spectral amplitude and was shown that the spectral slope of fluctuations is not changed. This opened the possibility of the statistical description of synchrotron emission for arbitrary $\gamma$\footnote{The numerical study in Herron et al. (2015) successfully tested the predictions in LP12 for fluctuating synchrotron intensities.}. Therefore, on the basis of this finding, we can say that, without Faraday rotation, the spectrum of polarized synchrotron emission should reveal the spectrum of the plane-of-the-sky magnetic field integrated along the LOS for any power law distribution of galactic cosmic ray electrons. However, it was not  clear how the observed spectrum of synchrotron polarization and underlying magnetic spectrum are related in the presence of Faraday rotation. This was the focus of the analytical study in Lazarian \& Pogosyan (2016; henceforth LP16). This study invoked a number of inevitable simplifying approximations and therefore it is important to test the LP16 predictions numerically.

With the ability to recover the statistics of the underlying magnetic field as well as underlying density of electrons it is possible to use the wealth of the synchrotron polarization data to understand the sources and sinks of turbulent energy in the Milky Way galaxy as well as for other galaxies and even clusters of galaxies. This research can help to constrain the driving and understand how energy is injected on large scales and transferred to smaller scales. Furthermore, our understanding of the spectrum can make it possible to produce a precise polarization map arising from magnetized turbulence and remove the synchrotron foreground in future CMB polarization observations.

We expect that high sensitivity of new generation telescopes, e.g., Square Kilometer Array (SKA) and LOw Frequency ARray (LOFAR), will carry numerous information to map synchrotron polarization (Beck \& Wielebinski 2013). For example, cosmic ray electrons with relatively low energies ($\sim$GeV) originated from supernova remnants in the Galactic disk generate synchrotron emission at low frequencies. LOFAR can detect their propagation and evolution process at low frequencies, which result in fluctuations of polarization, and their relation to the properties of turbulent magnetic field. 
Moreover, the observational facility, such as the VLA and Australian Square Kilometre Array Pathfinder (ASKAP), can produce considerable observation for making sensitive polarization images of the entire sky. Polarization Sky Survey of the Universe's Magnetism (POSSIM; Gaensler 2010) conducted with ASKAP is ongoing project on Faraday structure determination, which covers a frequency range from 1100MHz to 1400MHz (Sun et al 2015). 
In future we will be able to reconstruct the polarization spectra of synchrotron emission and Faraday rotation by comparing with those observations, and obtain detailed magnetic field statistics from polarization observations with SKA at both lower frequencies and higher frequencies in the Milky Way and intracluster medium.

In this paper, we investigate spectral behavior of polarized synchrotron fluctuations in the presence
of Faraday rotation. Since the effects of Faraday rotation is proportional to $\lambda^2$, where $\lambda$ is the wavelength, we show how the spectrum changes as the wavelength increases. We describe numerical methods in Section \ref{sect:numeric}. We present results in Section \ref{sect:results}, discussions in Section \ref{sect:discussion}, and summary in Section \ref{sect:summary}. In Section \ref{sect:results}, we include calculations for interferometric observations.

\section{Numerical Methods} \label{sect:numeric}

\subsection{Numerical simulations}  \label{sect:simulation}

\begin{figure}[h]
\centering
\includegraphics[scale=.52,angle=0]{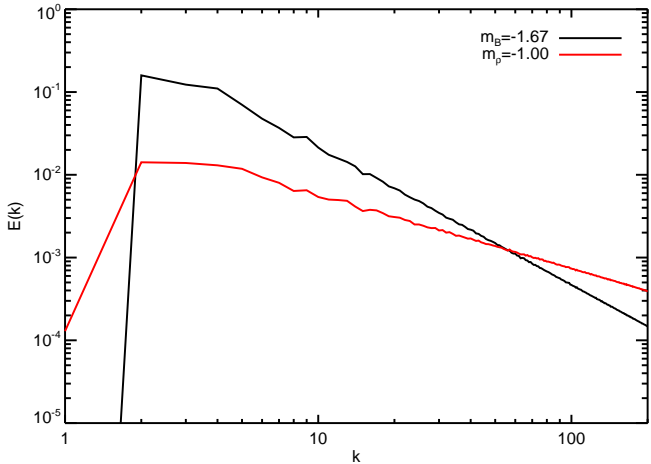}
\caption{Shell-integrated 1D spectrum of magnetic field (black solid line) and density (red solid line). Since the mean magnetic field and the mean density are 1 in code units, the black solid line actually represents the spectrum of the product of uniform density and fluctuating magnetic field. Similarly, the red solid line denotes the spectrum of the product of fluctuating density and uniform magnetic field. Spectral index for magnetic field is $-5/3$, but that for density is $-1.00$. Note that, when the spectral index for 3D power spectrum is $-11/3$, that for shell-integrated 1D spectrum is $-5/3$.
}
\label{fig:syn_index}
\end{figure}

The use of synthetic data is the simplest approach to simulate turbulence. 
We generate 3D synthetic data cubes of magnetic field and electron density 
in a periodic box of size $2\pi$ in Cartesian coordinate system (x,y,z), where z is along the LOS. 
The numerical resolution is $512^3$ for all synthetic data.
In general, the magnetic field  ($\textbf{B}$) consists of a uniform background field ($\textbf{B}_0$) and a fluctuating field  ($\textbf{b}$).
But, for synthetic magnetic fields, we assume $\textbf{B}_0=0$, so that $\textbf{B}=\textbf{b}$.
The generation of the synthetic data is actually done in Fourier space:
\begin{equation}
\textbf{B}(\textbf{x})=\textbf{B}_{0}+\sum_{|k|=2}^{k_{max}-1}\textbf{A}(\textbf{k})e^{i\chi},
\end{equation}
where $k_{max}=512/2$, \textbf{A}(\textbf{k}) is a real vector that is perpendicular to the wavevector \textbf{k}, $\chi=\textbf{k}\cdot\textbf{x}+\phi\left(\textbf{k}\right)$, $\textbf{x}=(x,y,z)$, and $\phi\left(\textbf{k}\right)$ is a random number in the range of 0 and 2$\pi$. 
We enforce $\phi(-\textbf{k})=-\phi(\textbf{k})$ for reality condition. 
The amplitude of each Fourier component randomly fluctuates, but on average the amplitude follows $|\textbf{A}(\textbf{k})|^2=\text{C}k^{m}$, where m is a constant $(e.g.,~\text{m}=-11/3 ~\text{for Kolmogorov spectrum})$ and C is a normalization constant that makes the r.m.s. magnetic field fluctuation of order unity. Fluctuations of magnetic field reflect the power-law spectra of the underlying magnetohydrodynamic (MHD) turbulence (Cho \& Lazarian 2010). 
Since it is the best-known spectrum for turbulence, we start with a Kolmogorov spectrum but we also use other spectral indices.  For magnetic field and density the peak of the spectrum appears at $k=2$. Density consists of a uniform density and a fluctuating density. The uniform density is set to one and the r.m.s. density is of order unity. When the power law index for density spectrum is -11/3, the product of the uniform density $(\rho_{0})$ and the fluctuating magnetic field $(b)$ is slightly larger than that of the uniform magnetic field $(B_{0})$ and the fluctuating density $(\delta \rho)$, i.e., $\rho_0 b > B_0 \delta \rho$. On the other hand, when the power law index for density spectrum is -3, $\rho_0 b$ is still larger than $B_0 \delta \rho$ on large scales. However, since the density spectrum is shallower than the magnetic spectrum, the fluctuation of density decreases relatively slowly as the scale decreases. As a result, $\rho_0 b$ becomes smaller than $B_0 \delta \rho$ on small scales.  The transition happens at near $k\sim60$ (see Figure  \ref{fig:syn_index}). In some runs, we only use the uniform density.

We also simulate ideal isotropic MHD turbulence using a code based on a third-order accurate hybrid essentially non-oscillatory (ENO) scheme in a periodic box of size $2\pi$:
\begin{eqnarray}
\partial\rho /\partial t + \triangledown\cdot(\rho\textbf{v})=0, \nonumber\\
\partial\textbf{v}/\partial t + \textbf{v}\cdot\triangledown\textbf{v}+\rho^{-1}\triangledown(a^{2}\rho)-(\triangledown\times\textbf{B})\times\textbf{B}/4\pi\rho =\textbf{f} \nonumber\\
\partial\textbf{B}/\partial t -\triangledown\times(\textbf{v}\times\textbf{B})=0, \nonumber
\end{eqnarray}
with $\triangledown\cdot\textbf{B}=0$, where $\rho$ is density, a is the sound speed, $\textbf{f}$ is a random driving vector, and other variables have their usual meanings. The r.m.s. velocity is maintained to be approximately unity, so that \textbf{v} can be viewed as the velocity measured in units of the r.m.s. velocity of the system. The simulation was performed with a resolution of $512^{3}$ grid points. We drive turbulence solenoidally in Fourier space. Both the Alfven mach number \textbf{($M_{A}=v_{rms}/V_{A}$}, where \textbf{$v_{rms}$} is the r.m.s. velocity and $V_{A}$ is the Alfven speed) and the sonic Mach number \textbf{($M_s =v_{rms}/a$)} of the turbulence are approximately 0.7. For the calculation of synchrotron polarization (in Section \ref{sect:noise_mhd}), we take magnetic field directly from the simulation data and assume the electron number density is proportional to $\rho$. In actual calculations, we use $n_e = \rho$.

\subsection{Polarization from synchrotron radiation}

The magnetic field, \textbf{B}, and the electron energy distribution determine synchrotron emission. We assume isotropic pitch angle distribution and a power-law energy distribution of electron population characterized by  the power-law index of electron energy distribution, \textit{p}:
\begin{equation}
N(E)dE=N_{0}E^{-p}dE,
\end{equation}
where $E$ is the electron energy, $N$ is the number of electrons per $E$ per unit volume, and $N_{0}$ represents homogeneous distribution of relativistic electrons.

The combination of the Stokes parameters provides a valuable description of synchrotron polarization.
In this paper, we focus only on linear polarization defined by the Stokes parameters Q and U:
\begin{equation}
P\equiv Q+iU.
\end{equation}

We can write polarized intensity observed at a two-dimensional position \textbf{X} on the plane of the sky at wavelength $\lambda$ as follows:
\begin{equation}
\text{P}\left(\textbf{X},\lambda ^{2}\right)=\int_{0}^{L}dz P_{j}(\textbf{X},z)e^{2i\lambda^{2}\Phi(\textbf{X},z)}
\end{equation}
where  $\textbf{X}=(x,y)$, $\textit{P}_{j}$ is the intrinsic polarized intensity density, L is the extent of the source along the LOS, and the exponential factor describes Faraday rotation from the source point at z along the line of sight to the observer.
The Faraday rotation measure, $\Phi(\textbf{X},z)$ is given by
\begin{equation} \label{eq:RM}
\Phi(\textbf{X},z)= \int_{0}^{z} \left(\frac{n_{e}(z)}{0.01cm^{3}}\right) \left(\frac{B_{z}(z)}{1.23\mu G}\right) \left(\frac{dz}{100pc}\right) \quad \text{rad}\,\text{m$^{-2}$}
\end{equation}
where $\textit{n}_{e}$ is the number density of electrons, and $B_{z}$ is the strength of the parallel component to the LOS component of the magnetic field. Here, we assume that the rotation measure is only attributed to the source region but not any intervening material between the observer and the source region along the LOS. This normalization is equivalent to the assumption that polarized synchrotron radiation is emitted from a region (e.g., Galactic halo) which has a size of 100pc and is located at 1kpc from the observer, electron number density in this region is $0.01cm^{-3}$, and magnitude of magnetic field is 1.23$\mu$G. Using this normalization, we can convert polarization in computational units (Section \ref{sect:test}, \ref{sect:dpdl2}, and \ref{sect:resolution}) to real units (Section \ref{sect:interfero}).

\subsection{Dataset for interferometer} \label{sect:datainterfero}
The interferometric observational data can be directly used to get spectra of turbulence (see LP16). 
In Sections \ref{sect:resolution} and \ref{sect:interfero} we simulate realistic interferometric observations with a finite beam size and a noise.
The procedure of simulating an interferometric observation that we employ is as follows:
\begin{enumerate}
\item Using the whole 3D magnetic field and density data, we calculate synchrotron polarization map projected on the plane of the sky for all LOSs.
\item To reflect a finite beam size of telescopes, we smooth the map using a Gaussian kernel:
         \begin{equation}
          g(x,y) \propto exp( -(x^2+y^2)/2{\sigma_{beam}}^{2}), 
         \end{equation}
         where $\sigma_{beam}$ ($=\theta_{FWHM}/2\ln2)$ is the resolution of the telescopes.
\item We obtain the complete 2D power spectrum of the smoothed polarization map.
\item To mimic an interferometric observation, we randomly select the wave-vectors corresponding to the baselines of a telescope array in Fourier space. 
\item We identify the 2D power spectrum at each selected wave-vector, \textbf{K}.
\item We  add a Gaussian random noise to the 2D power spectrum at each selected wave-vector. The amplitude of the noise is independent of the wavenumber.
\end{enumerate}

\section{Results} \label{sect:results}
\subsection{Statistics: Power Spectrum} \label{sect:test}
\subsubsection{Statistics without Faraday rotation}

\begin{figure*}[th]
\includegraphics[scale=.52,angle=0]{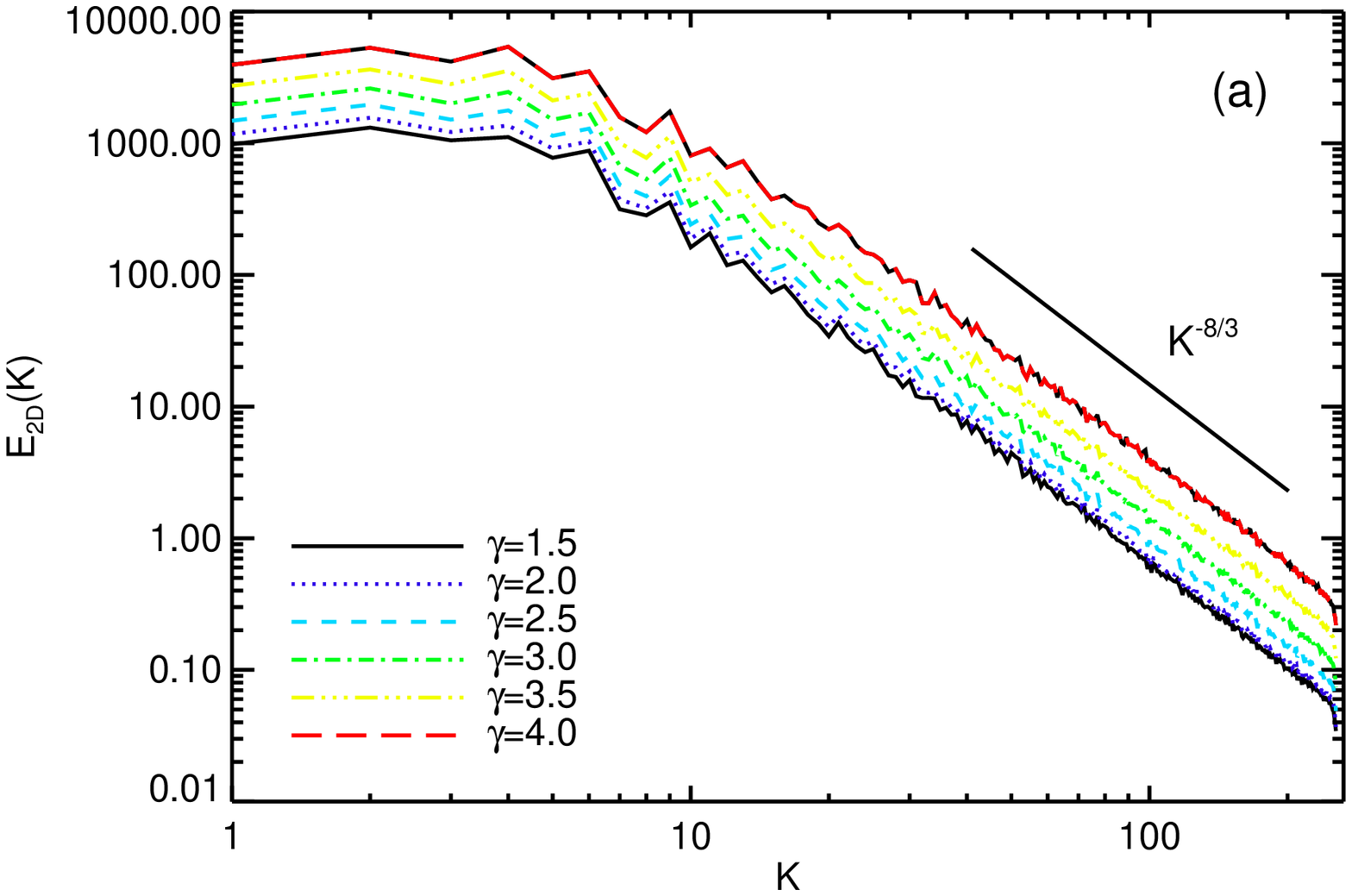} 
\includegraphics[scale=.52,angle=0]{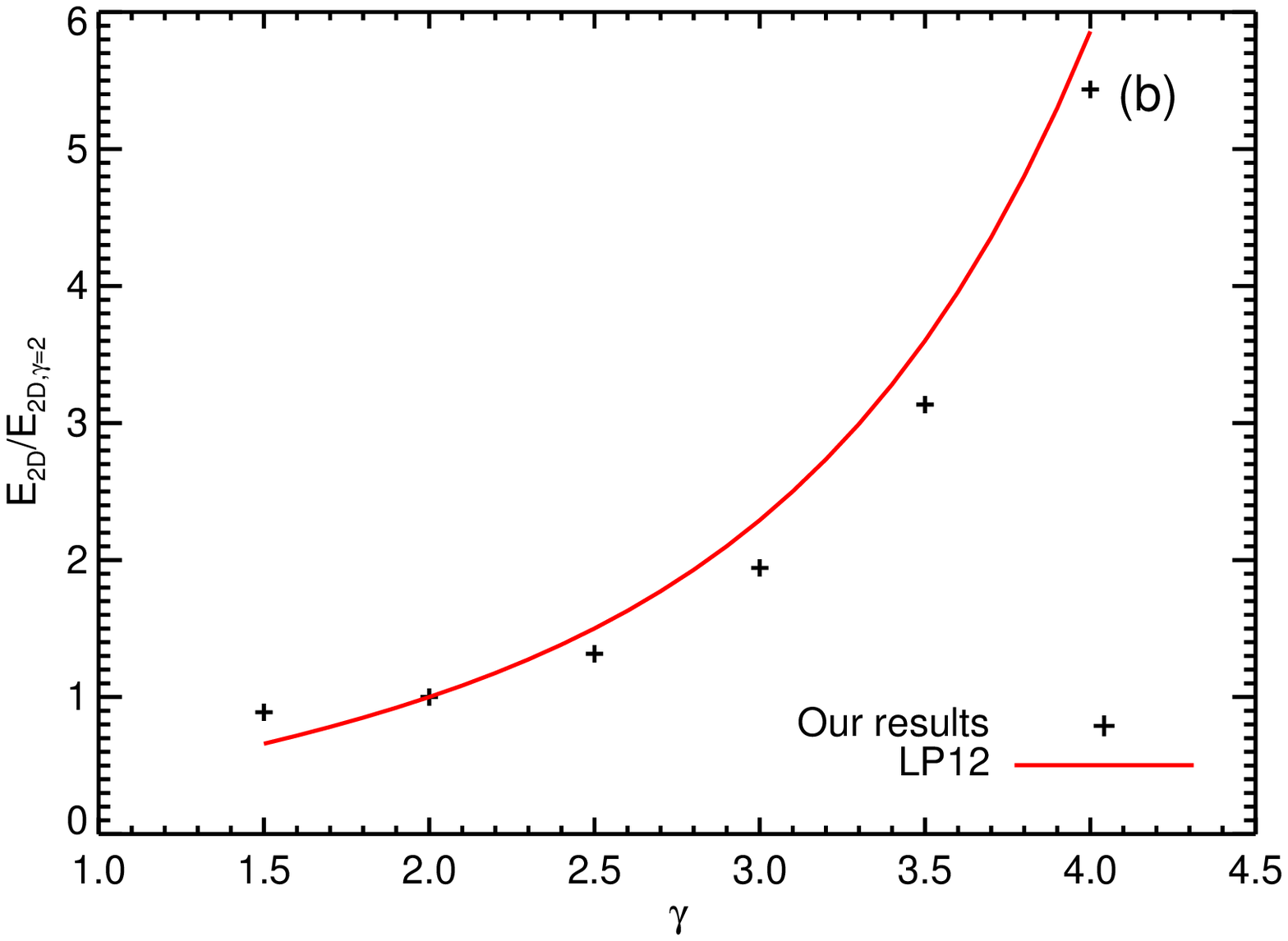}
\caption{Effects of the power-law index $\gamma$. Here, the polarized synchrotron emissivity depends on $B^{\gamma}_{\perp}$ and $\gamma=(p+1)/2$, where $p $ is the power-law index of electron energy distribution. (a) The 1D spectrum of polarized synchrotron emission for $1.5 \leq\gamma \leq 4$. (b) The comparison between the measured amplitudes of polarized synchrotron emission normalized by $E_{2D,\gamma=2}$ and the theoretical prediction in LP12.
}
\label{fig:syn}
\end{figure*}

The synchrotron emissivity depends on $\textbf{B}_{\perp}^{\gamma}$, where $\textbf{B}_{\perp}$ is the plane-of-the-sky magnetic field. It was predicted in LP12 that the variations of the spectral index ($p$) of relativistic electron energy distribution change the amplitude of the fluctuations, but not the spectral slope of the synchrotron power spectrum. The polarized synchrotron emissivity also depends on $\textbf{B}_{\perp}^{\gamma}$. Therefore we expect a similar behavior for polarized synchrotron emission. Below we test this.

The spectrum of polarized synchrotron radiation moves upward as $\gamma$ increases as shown in Figure \ref{fig:syn}(a). We explore the dependence of the synchrotron polarization statistics, i.e., power spectrum, on the power-law index $\gamma$. We consider $\gamma$'s ranging from 1.5 to 4, which cover all possible important cases of astrophysical power-law indices. To see the effect of $\gamma$, we fix the spectrum of turbulence: we use Kolmogorov spectra for magnetic field and density. As we can see in Figure \ref{fig:syn}(a), the spectra exhibit extended power-laws at large wavenumbers for all values of $\gamma$ shown in the figure. The measured power-law slopes  at large wavenumbers are very close to that of a Kolmogorov spectrum, i.e.,~$-8/3$ for \textit{ring-integrated} 1D spectrum $E_{2D}(K)$ (compare the spectra with the black solid straight line in the figure). We can also see that the compensated spectra have plateaus at large wavenumbers, which means that the \textit{ring-integrated} 1D spectra $E_{2D}(K)$ at large wavenumbers are proportional to $K^{-8/3}$.

To see the dependence of the spectrum on $\gamma$ more quantitatively, we plot the amplitude of polarized synchrotron emission spectra for different spectral indices in Figure \ref{fig:syn}(b). The red solid line is obtained from the Equation (37) in LP12 that show the dependence of amplitude on spectral index, and the black plus sings represent measured amplitudes of polarized synchrotron emission normalized by $E_{2D,\gamma=2}$. We can clearly see that the measured spectral amplitude is in agreement with the prediction of LP12.

\subsubsection{Effects of Faraday rotation}

\begin{figure*}[h]
\centering
\includegraphics[scale=.52,angle=0]{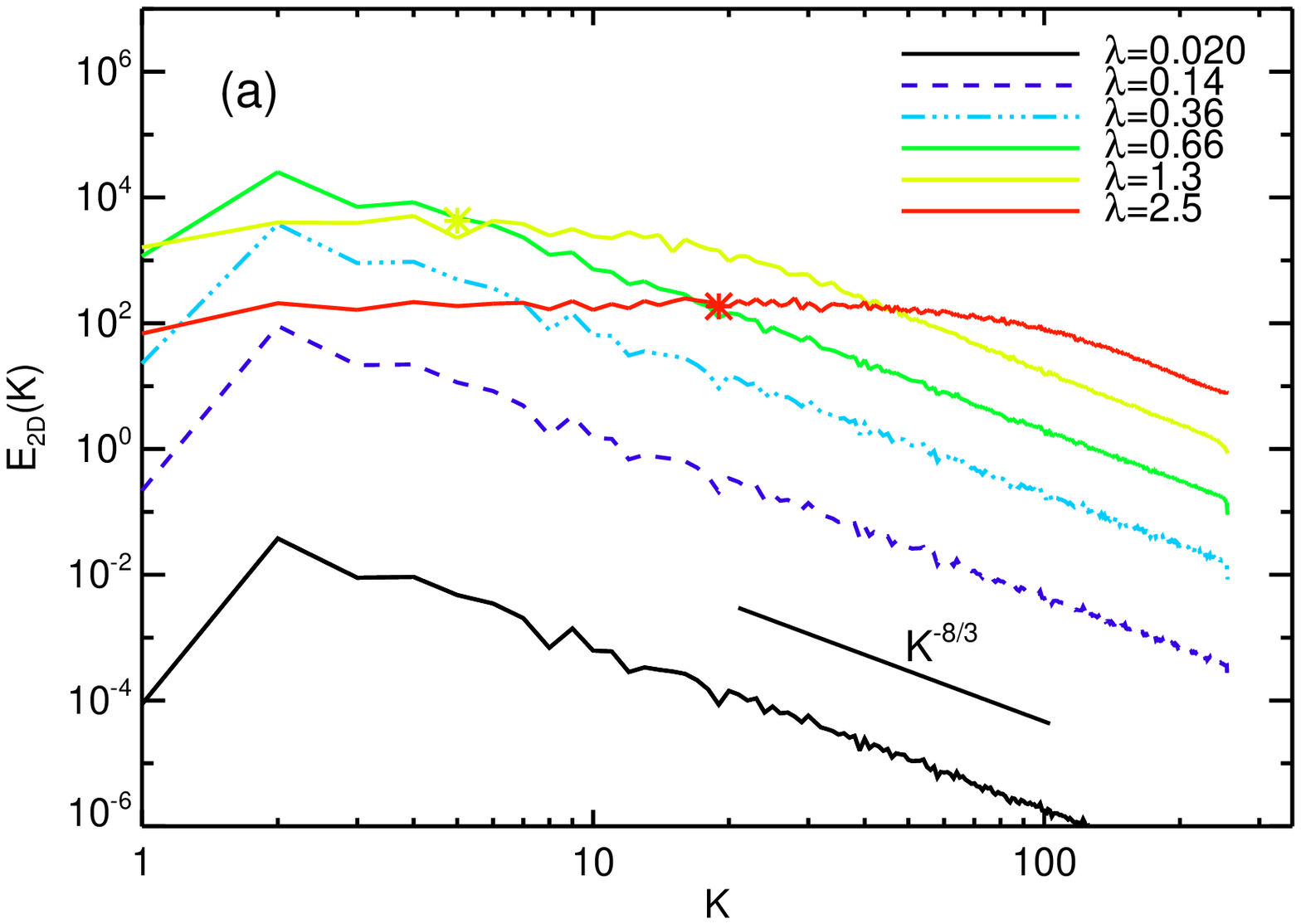}
\includegraphics[scale=.52,angle=0]{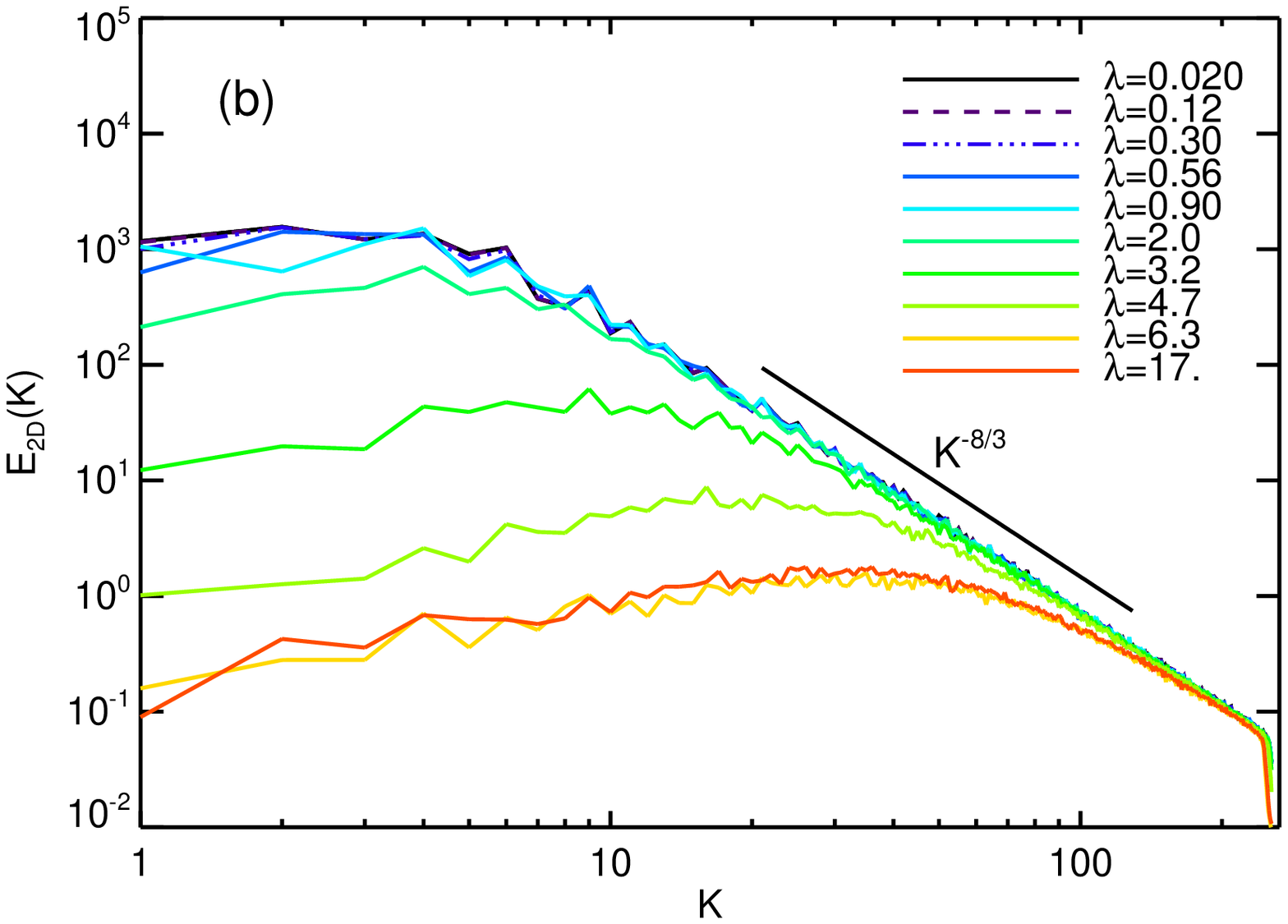}
\caption{(a) \textit{Ring-integrated} 1D spectrum $E_{2D}(K)$ arising only from fluctuations of Faraday rotation at different wavelengths in code units. We use fluctuating electron number density and $\textbf{B}_{\parallel}$, but uniform intrinsic polarized emission. The asterisks mark the wavenumber $K=2\pi \lambda^{2}\left<n_{e}\left| \textbf{B}_{\parallel}\right|\right>$. When K is smaller than this wavenumber, Faraday depolarization is significant. (b) \textit{Ring-integrated} 1D spectrum $E_{2D}(K)$ arising only from fluctuations of synchrotron emission. We use fluctuating plane-of-the-sky magnetic field, but uniform electron density and magnetic field along the LOS.
}
\label{fig:FR}
\end{figure*}

Faraday rotation depends on wavelength as well as the electron number density and the strength of magnetic field parallel to the LOS (Equation \ref{eq:angle}). To see the effects of Faraday rotation only, we calculate synchrotron polarization with uniform intrinsic polarized emission. That is, we use $Q/I=1$ and $U/I=0$ for intrinsic synchrotron emission at all points in space and calculate Faraday rotation.
We, however, use fluctuating LOS magnetic field and electron number density. The average electron number density is set to one. 
We plot the results in Figure \ref{fig:FR}. In Figure \ref{fig:FR}(a), we can see that the spectrum goes up as the wavelength increases, which means that the polarization deviates more and more from $Q/I=1$ and $U/I=0$ at longer wavelengths. When $\lambda \gtrsim 1$ the small-$K$ part of the spectrum no longer moves up with increasing wavelength, while the large-$K$ part of the spectrum continues to move up.

The reason why the small-$K$ part of the spectrum does not move up when $\lambda \gtrsim 1$ is Faraday depolarization effect. Figure \ref{fig:FR}(a) suggests that Faraday depolarization happens on the large scale first. In fact, Faraday depolarization is significant when $K \lesssim 2\pi \lambda^{2}\left<n_{e} \left| \textbf{B}_{\parallel}\right|\right>$, and it is insignificant when $K \gg 2\pi \lambda^{2}\left<n_{e} \left| \textbf{B}_{\parallel}\right|\right>$ (in the synthetic data, $\left<n_{e} \left|\textbf{B}_{\parallel}\right|\right> \sim 0.5$)\footnote{In this paper, we assume that the size of the computational domain is $2\pi$. If the size of the system is L, then the expression becomes $K \lesssim L \lambda^{2}\left<n_{e} \left| \textbf{B}_{\parallel}\right|\right>$, where K is the number of waves that exist over the size L.}. The asterisks in Figure \ref{fig:FR}(a) denote $K=2\pi \lambda^{2}\left<n_{e} \left| \textbf{B}_{\parallel}\right|\right>$. As the wavelength increases, the scale above which the depolarization effect is significant becomes smaller and smaller. When $\lambda^{2} \sim \frac{K_{max}}{2\pi\left<n_{e} \left| \textbf{B}_{\parallel}\right|\right>}$, the polarized emission from each grid point becomes completely uncorrelated and the emission from each grid point contributes randomly\footnote{When this happens, all Fourier modes have similar powers. In this case, the ring-integrated energy spectrum, $E_{2D}(K)$, becomes proportional to the number of Fourier modes in $K-0.5 \leq |\textbf{K}|<K+0.5$, which makes the spectrum proportional to K.}, which makes the spectrum proportional to $K$.

In the previous example in this paragraph, we calculated Faraday rotation, assuming a constant intrinsic synchrotron polarization. When we use a turbulence spectrum for the plane-of-the-sky magnetic field, which are responsible for synchrotron emission, we may be able to see the effect of Faraday depolarization more clearly. For simplicity, let us assume that the density and the LOS component of magnetic field are constants. This is definitely a toy model to probe the effects of Faraday rotation induced by the simplest realization of the Faraday rotation effect. In Figure \ref{fig:FR}(b), we present the resulting spectrum of synchrotron polarization, in which we use a Kolmogorov spectrum for the plane-of-the-sky magnetic field. When $\lambda$ is very small (e.g., see the black solid curve), the Faraday rotation is small and the spectrum does not suffer from the Faraday depolarization effect. As $\lambda$ increases, the spectrum at small $K$ decreases due to the Faraday depolarization effect. The fact that the spectrum at large $K$ does not show much change implies that the depolarization effect is negligible at large $K$.

A similar effect may be present also in the case when the Faraday rotation effect is induced by a regular magnetic field $\textbf{B}_0$ and fluctuations of density. In this case we may also see distortion of the spectrum at small $K$ and the preservation of the spectral slope at large $K$. The effect of additional fluctuations arising from the Faraday fluctuations may increase the amplitude of the measured spectrum.

Dependence of spectrum on wavelength is very different for Faraday rotation fluctuations (Figure \ref{fig:FR}(a)) and for synchrotron polarization fluctuations (Figure \ref{fig:FR}(b)). When Faraday rotation fluctuations are much larger than synchrotron polarization fluctuations, the spectrum will show change with wavelength as plotted in Figure \ref{fig:FR}(a). When synchrotron polarization fluctuations are dominant, the spectrum will be changed as wavelength increases as plotted in Figure \ref{fig:FR}(b). We can use these behaviors to distinguish two cases.

\subsubsection{Polarization by synchrotron radiation and Faraday rotation}

\begin{figure*}[th]
\centering
\includegraphics[scale=.52,angle=0]{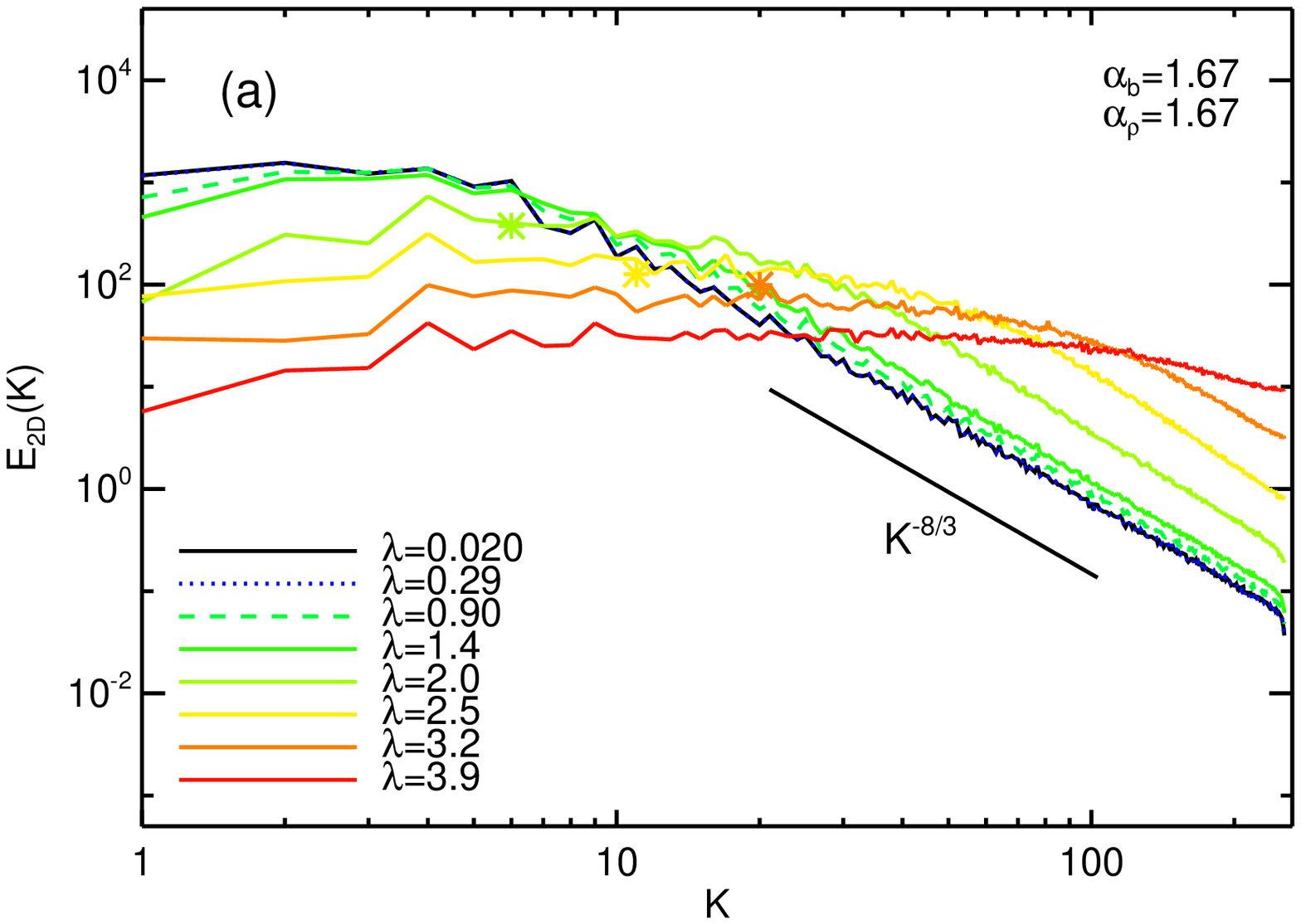} 
\includegraphics[scale=.52,angle=0]{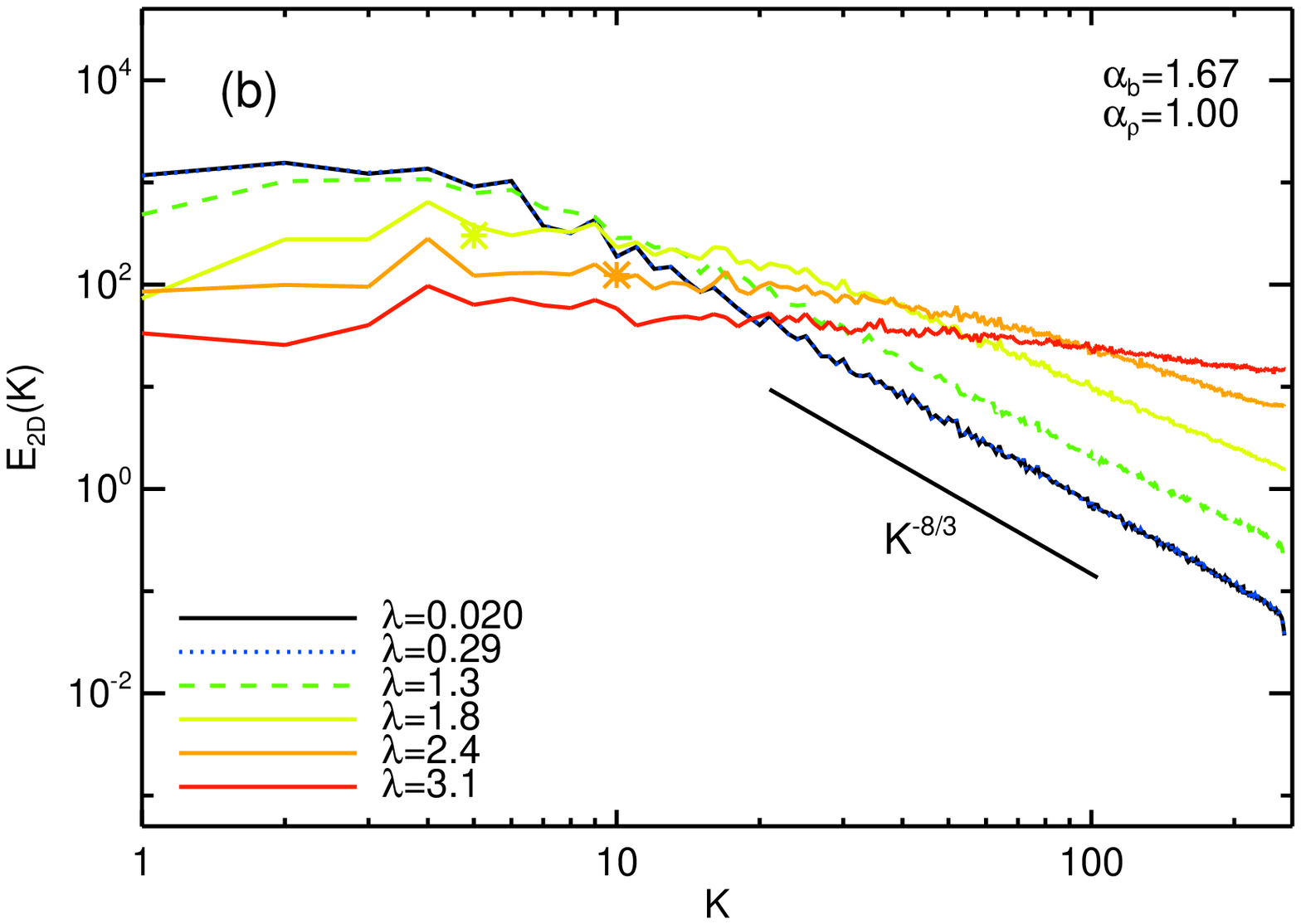}
\caption{\textit{Ring-integrated} 1D spectrum $E_{2D}(K)$ arising from fluctuations of synchrotron radiation and Faraday rotation. We use fluctuating electron number density and fluctuating \textbf{B} of synthetic data to calculate spectrum of polarized synchrotron emission. (a) The 3D power spectrum $P_{3D}(k)$ is proportional to $k^{-11/3}$ for both electron number density and \textbf{B}. (b) The 3D power spectrum $P_{3D}(k)$ is proportional to $k^{-9/3}$ for electron number density and $k^{-11/3}$ for \textbf{B}. Different curves correspond to different wavelengths.  The asterisks mark the wavenumber $K=2\pi \lambda^{2}\left<n_{e}\left| \textbf{B}_{\parallel}\right|\right>$. When K is smaller than this wavenumber, Faraday depolarization is significant.
}
\label{fig:synFR}
\end{figure*}

From Equation \ref{eq:RM}, we expect that fluctuations of electron density also introduce fluctuations of Faraday rotation, which will change the shape of the spectrum at long wavelengths. In this subsubsection, we take into account fluctuations of electron density as well as magnetic field and calculate synchrotron polarization.

We first assume that both density and magnetic field follow Kolmogorov spectra. Figure \ref{fig:synFR}(a) shows the resulting spectrum of polarization. When the wavelength $\lambda$ is very small (e.g., see the black solid curve), Faraday rotation is small and the spectrum reflects that of pure polarized synchrotron emission. As $\lambda$ increases, the effect of Faraday rotation begins to change the shape of the spectrum. It is interesting that not only the small-$K$ part but also the large-$K$ part of the spectrum is affected by Faraday rotation. The small-$K$ part of the spectrum goes down due to the depolarization effect. The rise of the spectrum at large-$K$ might be due to additional fluctuations in Faraday rotation measure arising from fluctuations in electron number density. According to Figure \ref{fig:synFR}(a), the power-law slope of the polarization spectrum does not change much with increased wavelength at large $K$. As $\lambda$ increases, Faraday depolarization becomes more and more important and the portion of the spectrum that shows a power-law dependence on $K$ gets limited. Therefore, if we are interested in measuring the power-law slope, we would better observe synchrotron polarization at small wavelengths.

In the case of Figure \ref{fig:synFR}(a), we have assumed that both the density and the magnetic field have the same spectral indices, i.e.,~both of them have Kolmogorov spectra. However, they can have different spectral slopes in real astrophysical environments. Therefore, we perform a calculation in which the density and the magnetic field have different spectral slopes. To be specific, we assume that the 3D spectral index of the magnetic field is $-11/3$ (i.e.,~the same as Kolmogorov) and that of density is $-3$ (i.e.,~shallower than Kolmogorov) (see Figure \ref{fig:syn_index}). Figure \ref{fig:synFR}(b) shows the resulting spectra of polarization at different wavelengths. When the wavelength $\lambda$ is very small (e.g., see the black solid curve), Faraday rotation is small and the spectrum reflects that of pure polarized synchrotron emission. The spectral slope at large wavenumbers is compatible with -8/3. As $\lambda$ increases, the effect of Faraday rotation begins to change the shape of the spectrum. The power at small wavenumbers goes down due to the Faraday depolarization effect, and the power at large wavenumbers goes up due to increased fluctuations in Faraday rotation. The spectral slope at large wavenumbers becomes shallower as $\lambda$ increases, implying that shallow density spectrum influences the small scale polarization spectrum. The overall behavior of the spectra is in agreement with the prediction in LP16.

\subsection{Spectrum of the polarization derivative with respect to $\lambda^{2}$} \label{sect:dpdl2}

\begin{figure*}[h]
\includegraphics[scale=.52,angle=0]{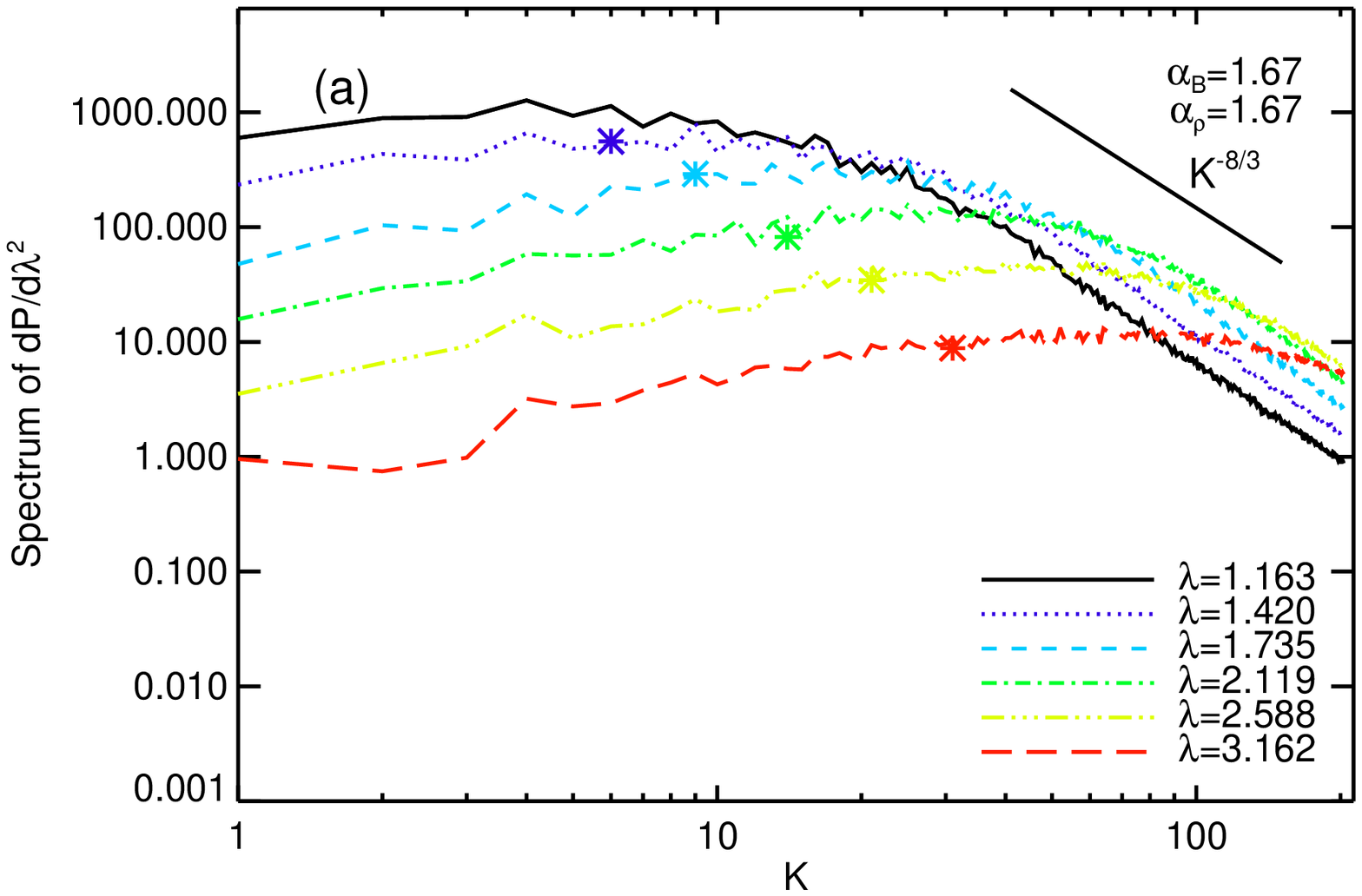}
\includegraphics[scale=.52,angle=0]{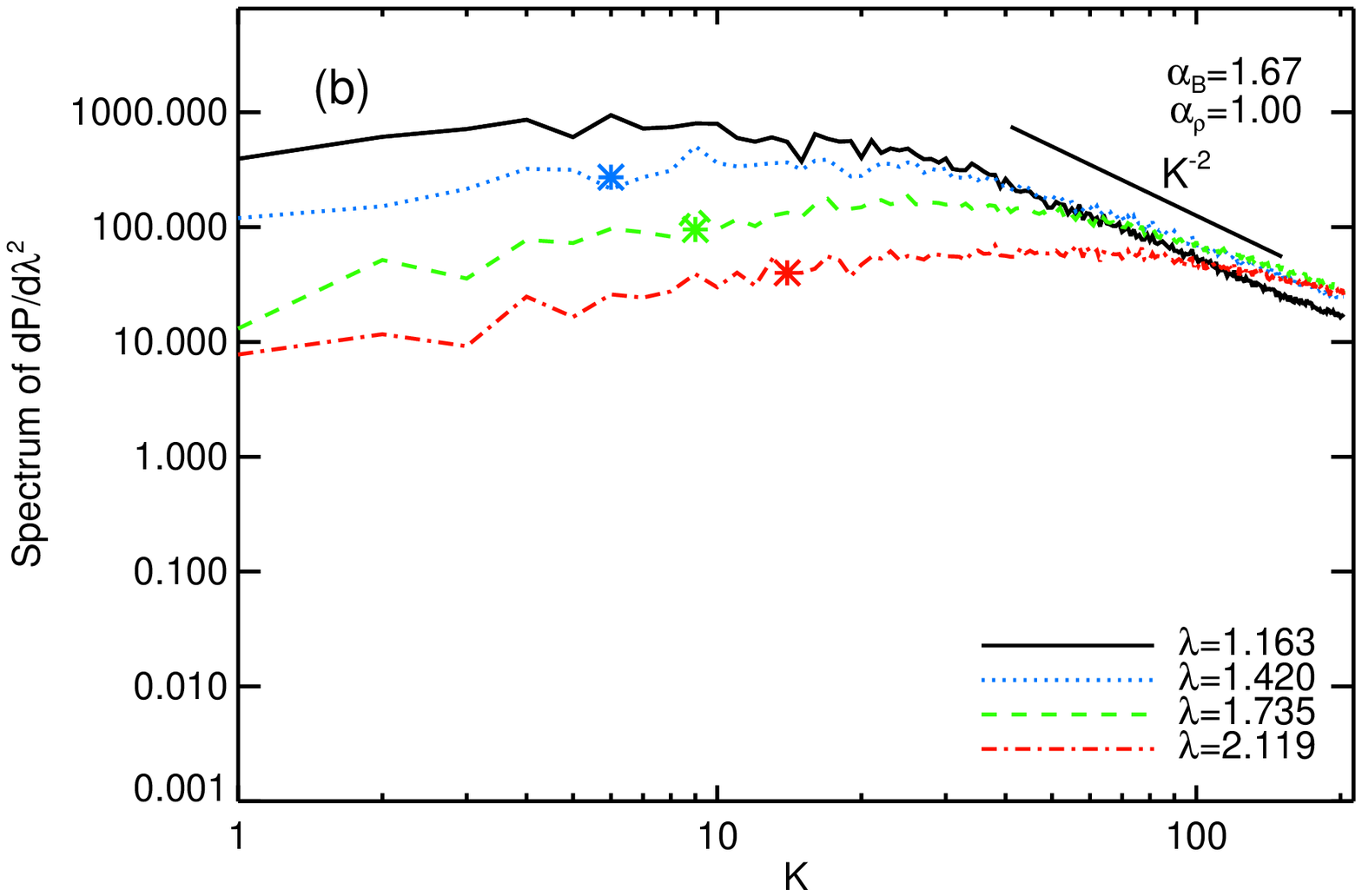}
\caption{Spectra of $dP/d \lambda^2$ for synthetic data. (a) Both magnetic field and density follow Kolmogorov spectra (i.e., the 1D spectrum, $E_{3D}(k) \propto k^{-5/3}$).  (b) The magnetic field follows a Kolmogorov spectrum, but the density follows a shallow spectrum (i.e., $E_{3D}(k) \propto k^{-1}$ for density). The asterisks mark the wavenumber $K=2\pi \lambda^{2}\left<n_{e}\left| \textbf{B}_{\parallel}\right|\right>$. When K is smaller than this wavenumber, Faraday depolarization is significant.
}
\label{fig:dpdl}
\end{figure*}

It was suggested in LP16 that the spectrum of $dP/d \lambda^2$ is also useful to recover the statistics of Faraday rotation. In this paper, we assume that the mean magnetic field is zero and test the suggestion in LP16.

We calculate the spectrum of $dP/d \lambda^2$ from the following procedure. First, using the synthetic data, we obtain maps of $P(\lambda_1)$ and $P(\lambda_2)$, where $|\lambda_1-\lambda_2|$ is small. Second, we calculate the quantity
\begin{equation}
         \frac{   P(\lambda_1) -P(\lambda_2) }{ \lambda_1^2 -\lambda_2^2 }.  \label{eq:dp_dl2}
\end{equation}
Third, we obtain the spectrum of the quantity.

We first consider the case in which both density and magnetic field have Kolmogorov spectra.
We calculate the spectra of the derivative and plot the results in Figure \ref{fig:dpdl}(a). 
When $\lambda$ is very small (e.g.,~black solid curve), the spectrum exhibits an extended power-law at large wavenumbers.
As $\lambda$ increases, the spectrum goes down at small wavenumbers and goes up at large wavenumbers.
As we can see in the figure, the spectral slopes for large wavenumbers are close to ${-8/3}$, which is the same as that of $P(\lambda)$ itself. This result is consistent with the prediction in LP16 (their Equation (105)).

We also consider the case in which the density follows a $k^{-3}$ 3D power spectrum and the magnetic field follows
a Kolmogorov spectrum (see Figure \ref{fig:syn_index}).
Figure \ref{fig:dpdl}(b) shows the resulting spectra.
The overall behavior of the spectra in Figure \ref{fig:dpdl}(b) is similar to that in Figure \ref{fig:dpdl}(a), but
the spectra at large $K$ values are much shallower than $K^{-8/3}$: it is a bit steeper but close to $K^{-2}$ in this case.

\subsection{Effect of telescope resolution}  \label{sect:resolution}

\begin{figure}[h]
\includegraphics[scale=.52,angle=0]{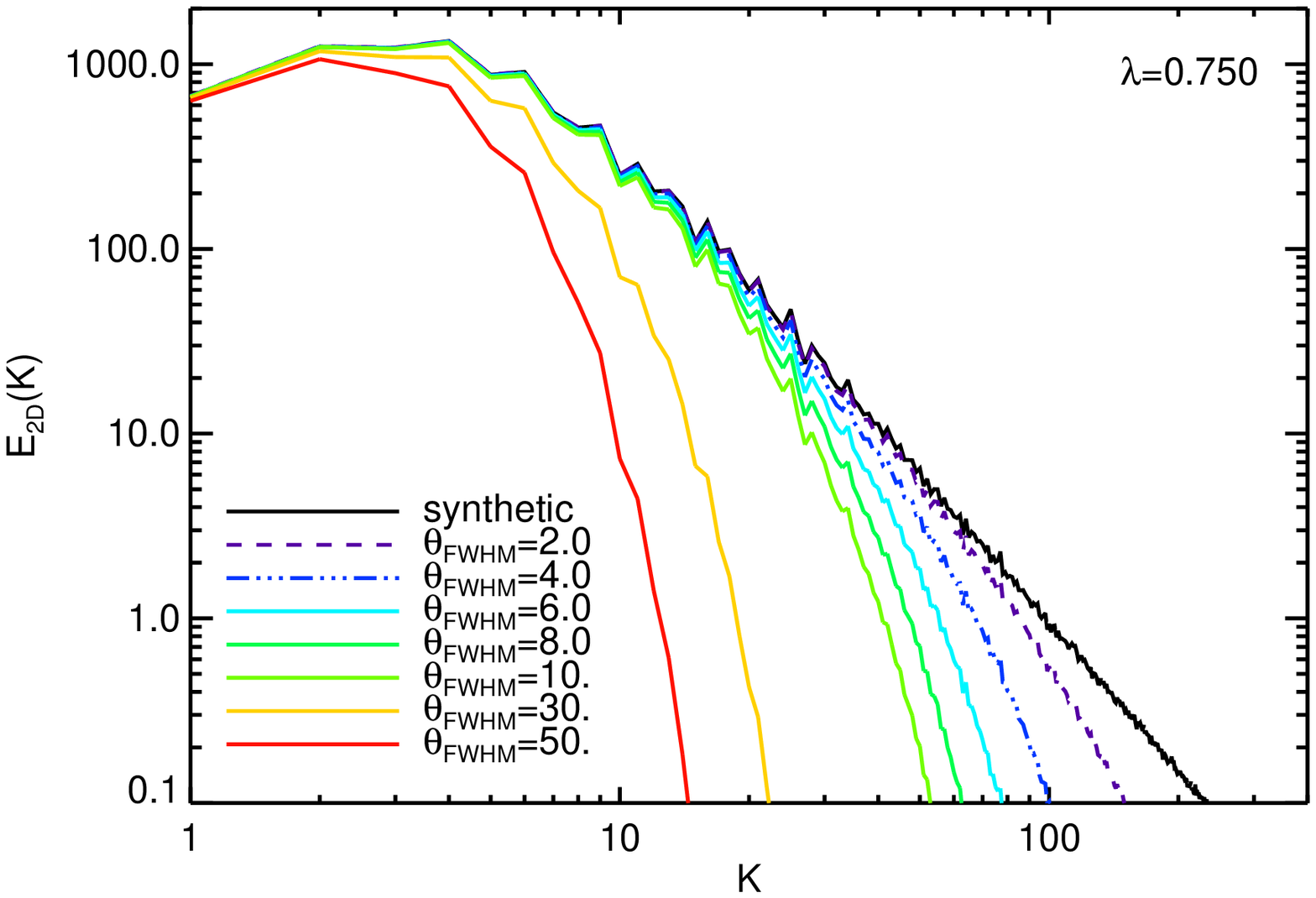}
\caption{The \textit{ring-integrated} 1D spectra $E_{2D}(K)$ of synchrotron polarization maps that reflect finite telescope resolutions. To mimic the effect of the telescope resolution, we smooth the maps with Gaussian kernels. The beam size ($\theta_{FWHM}$) is given in units of grid spacing. The observed wavelength is 0.75 in code units, at which Faraday rotation is small.
}
\label{fig:resolution}
\end{figure}

The resolution of telescopes affect the spectrum of synchrotron polarization because features smaller than the telescope resolution are smoothed out. To reflect a finite beam size of telescopes, we smooth the polarization map using a Gaussian kernel as described in Section \ref{sect:datainterfero}. In Figure \ref{fig:resolution}, we present power spectra of polarized emission smoothed with different telescope beam resolutions $\theta_{FWHM}$, which is given in units of grid points. The black solid curve represents the power spectrum without smoothing, which shows a well-defined power-law at large wavenumbers. As  $\theta_{FWHM}$ increases, the spectrum gradually deviates from the power-law. In fact, we expect the functional form of the spectrum to be proportional to 
\begin{equation}
   K^{m} \exp\left({-K^2/2 \sigma_K}\right),
\end{equation}
where $\sigma_K=1/\left(\sqrt{2 \pi} \sigma_{beam}\right)$.

As $\theta_{FWHM}$ becomes larger than $\sim$20, which corresponds to $\sim$1/25 of the size of the computational domain, the spectra 
show virtually no power-law inertial range. Obviously, it is advantageous to perform a high resolution observations to reveal the true power-law spectrum. Therefore, one should employ high-resolution observations with ground based interferometry to get the true power-law spectrum.

\subsection{Study of turbulence with interferometer}  \label{sect:interfero}

The advantage of the technique suggested in LP16 is that one can directly use interferometric data to get spectra of turbulence. Therefore there is no need for restoring the polarization image of the turbulent object. We numerically explore this possibility by using interferometric measurements {\it with just a few baselines}. The ability to measure synchrotron spectrum this way stems from the fact that the statistics of turbulence is rather simple object with high degree of symmetries (see more details in LP16). In what follows we take a few interferometric measurements of synthetic observations of the polarized synchrotron emission and add noise to them to simulate realistic observations. We explore how many baselines we require to recover the information about the underlying spectrum of turbulence. 

\subsubsection{Effect of the number of baselines}

\begin{figure*}[h]
\centering
\includegraphics[scale=.52,angle=0]{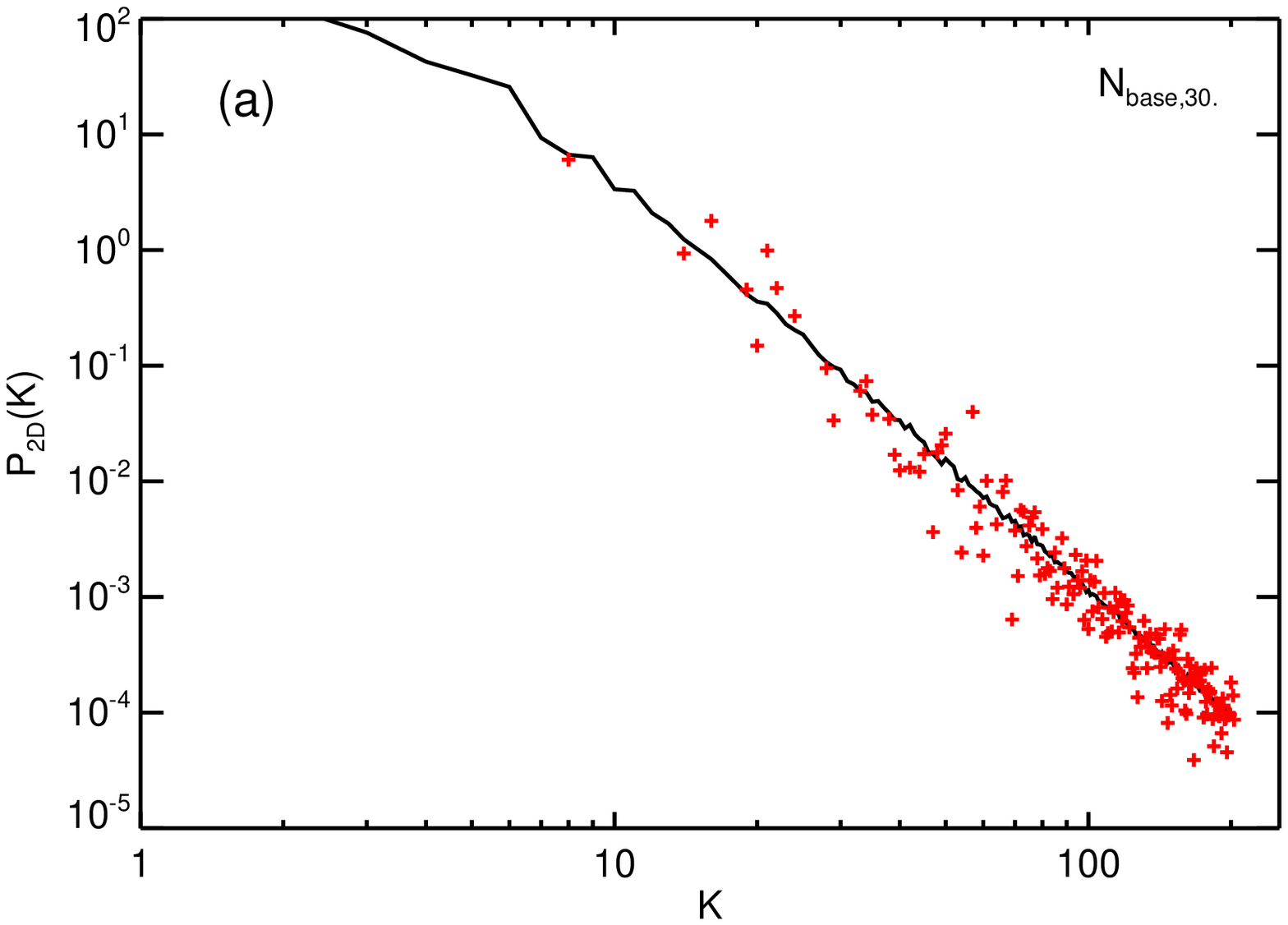}
\includegraphics[scale=.52,angle=0]{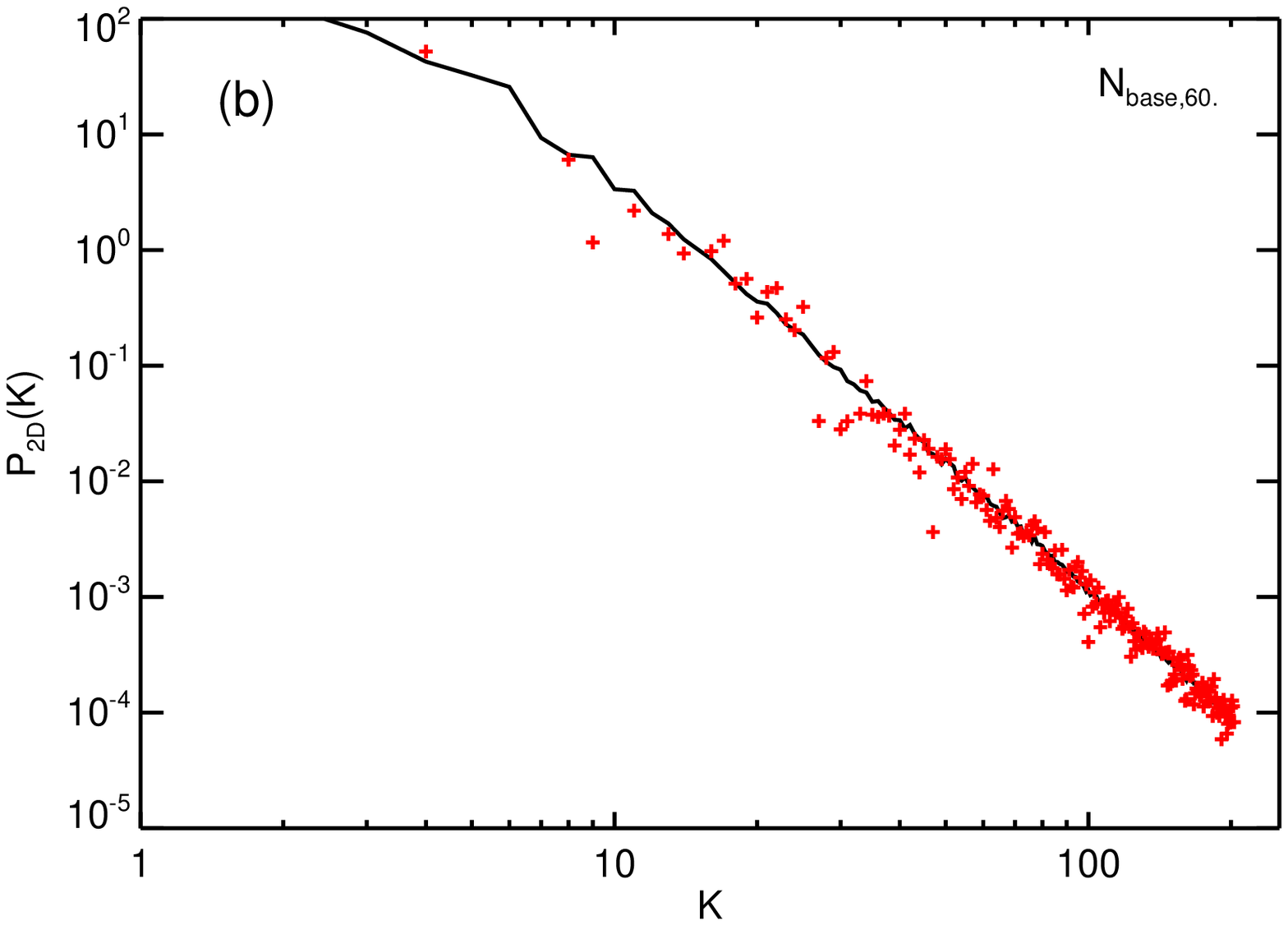}
\caption{The average 2D power spectrum from simulated interferometric observations with $N_{base}$ baselines ($plus$ symbols) and the true spectrum (the \textit{black solid} line). We assume that the beam size of telescopes is negligibly small: $\theta_{FWHM} \rightarrow 0$. (a) The number of baselines ($N_{base}$) is $30 \cdot 29/2~(\equiv N_{base,30})$. (b) $N_{base}=60 \cdot 59/2~(\equiv N_{base,60})$.
}
\label{fig:antena}
\end{figure*}

If we use interferometer that consists of $N$ telescopes for a short amount of time, we can obtain 2D power spectrum for $N_{base}=N(N-1)/2$ wave-vectors that correspond to the baselines. On the other hand, if we make use of Earth's rotation as in typical interferometric observations, we can increase the number of baselines $N_{base}$. Since we do not cover the whole wave-vector space, it may be difficult to recover the true power spectrum if $N_{base}$ is small. Then what will be a reasonable value for $N_{base}$? In Figure \ref{fig:antena}, we demonstrate that $N_{base,30}=30\times 29/2$ or $N_{base,60}=60\times 59/2$ is good enough to reconstruct the true power spectrum. The black solid curves correspond to the true power spectrum, which is calculated from complete
Fourier modes without any missing components. In order to mimic an interferometric observation with $N_{base}$ baselines, we randomly select $N_{base}$ wave-vectors and calculate average power spectrum based on the power at the selected wave-vectors:
\begin{equation} \label{eq:e2d}
P_{2D}\left(K \right)=\left<|\tilde{P}(\textbf{K})|^2 \right>= \sum_{i=1}^{N_K} |\tilde{P}(\textbf{K})|^2/N_K,   \label{eq:averagepower}
\end{equation}
where the summation is taken over $K-0.5 < |\textbf{K}|<K+0.5$, $\tilde{P}(\textbf{K})$ is the Fourier transform of $P(\textbf{x})$, and $N_K$ is the number of observed wave-vectors in the range $K-0.5 < |\textbf{K}|<K+0.5$. The red symbols in Figure \ref{fig:antena} represent the simulated average 2D power spectrum. In Figure \ref{fig:antena} (a) and (b), we assume that $N_{base}=N_{base,30}$ and $N_{base,60}$, respectively. The simulated average spectrum for $N_{base,30}$ shows a larger scatter than that for $N_{base,60}$. Nevertheless, both $N_{base,30}$ and $N_{base,60}$ can reproduce the true spectrum reasonably well.

\subsubsection{Effects of noise and finite beam size: simulations with synthetic data} 

\begin{figure}[h]
\centering
\includegraphics[scale=.52,angle=0]{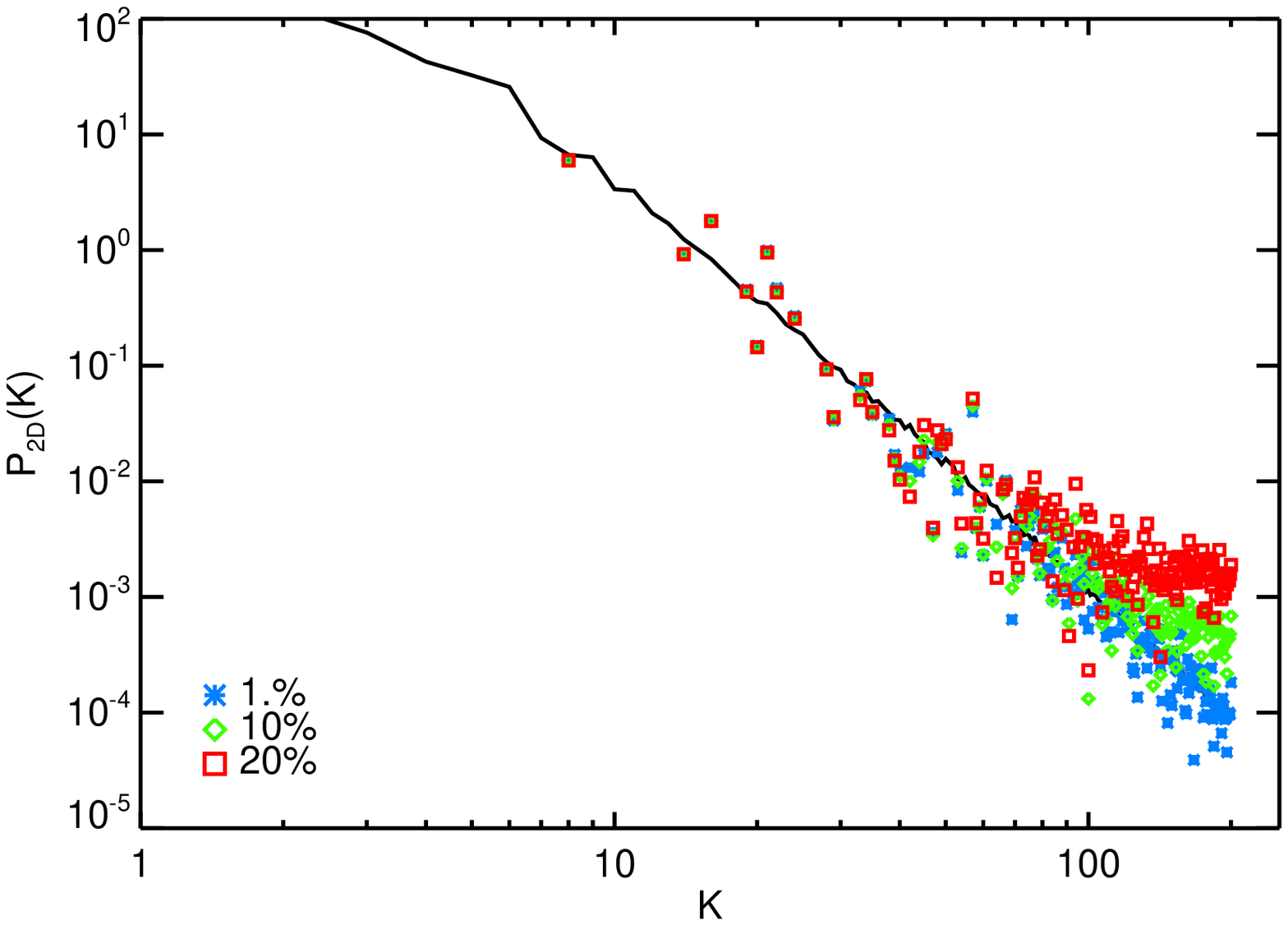}
\caption{The average 2D power spectrum from simulated interferometric observations with noise. Different symbols denote different noise levels: $1\%$ (blue $asterisks$), $10\%$ (green $diamonds$), and $20\%$ (red $squares$).
}
\label{fig:noise}
\end{figure}

First, we use synthetic data to see how the power-law slope is affected by noise. We follow the procedure of simulating a realistic interferometric observation described in Section \ref{sect:datainterfero}. We assume $N_{base}=30\cdot 29/2~(\equiv N_{base,30})$. Figure \ref{fig:noise} shows the 2D power spectra at $\lambda \sim 1$cm, at which the effect of Faraday rotation is non-negligible but still not significant. The black solid curve corresponds to the true spectrum without noise, which is calculated from the complete Fourier modes without any missing components. The blue $asterisk$, the green $diamond$, and the red $square$ symbols denote average 2D spectra from interferometric observations with random Gaussian noise. Different symbols represent different noise levels: the noise levels for the blue, the green, the red symbols are 1$\%$, 10$\%$, and 20$\%$ of the standard deviation of the true signal, respectively. The blue symbols follow the black solid curve quite well because the noise level is very low. Therefore, as long as the noise level is sufficiently low, an interferometric observation with $N_{base,30}$ baselines is good enough to recover the underlying turbulence spectrum. When the noise level increases, the power spectrum deviates from the black solid curve at large wavenumbers. The wavenumber at which the spectrum begins to deviate depends on the noise level: the turn-off wavenumber decreases as the noise level increases, which makes the power-law inertial range narrower.

\begin{figure*}[h]
\includegraphics[scale=.52,angle=0]{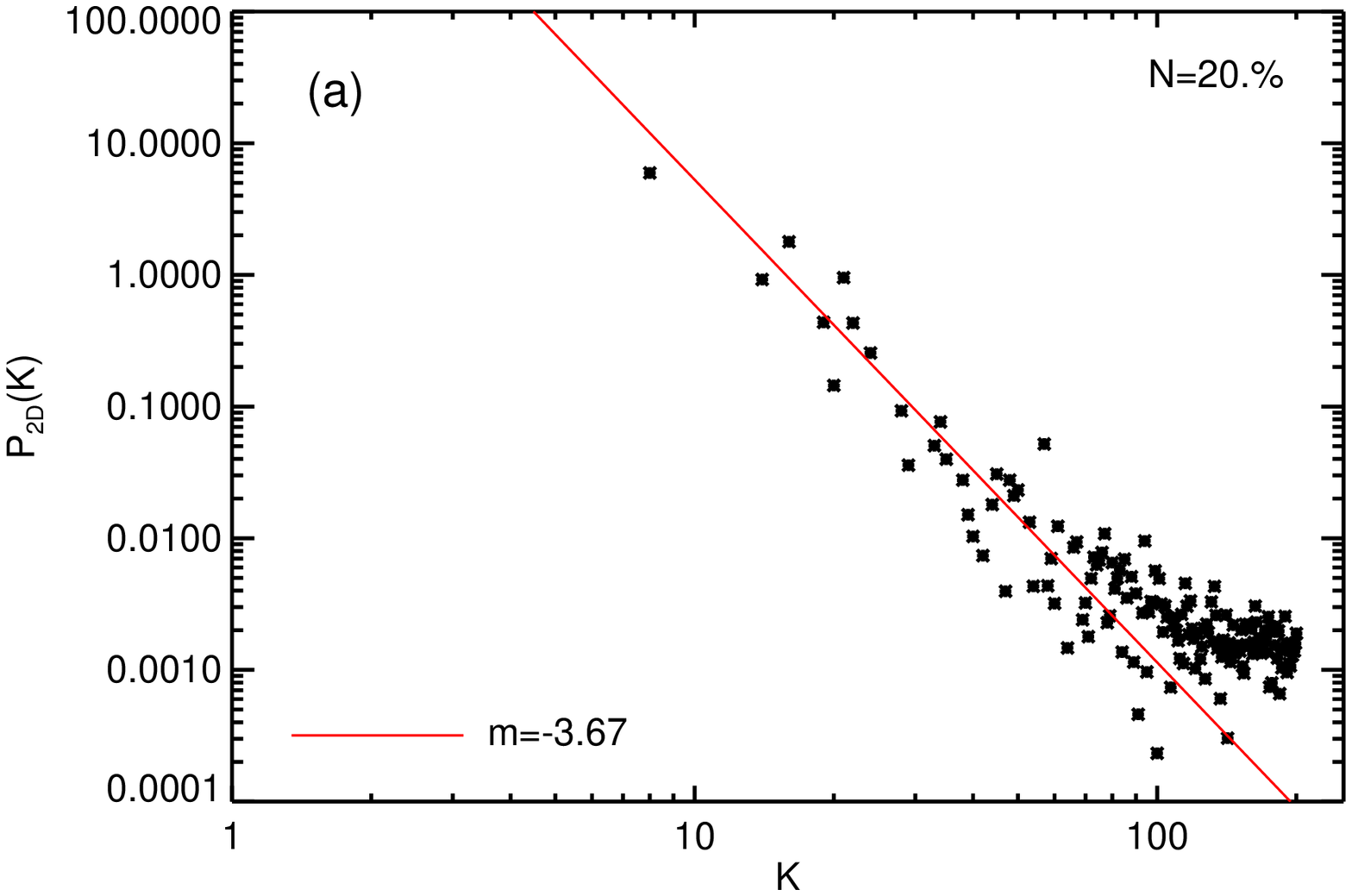}
\includegraphics[scale=.52,angle=0]{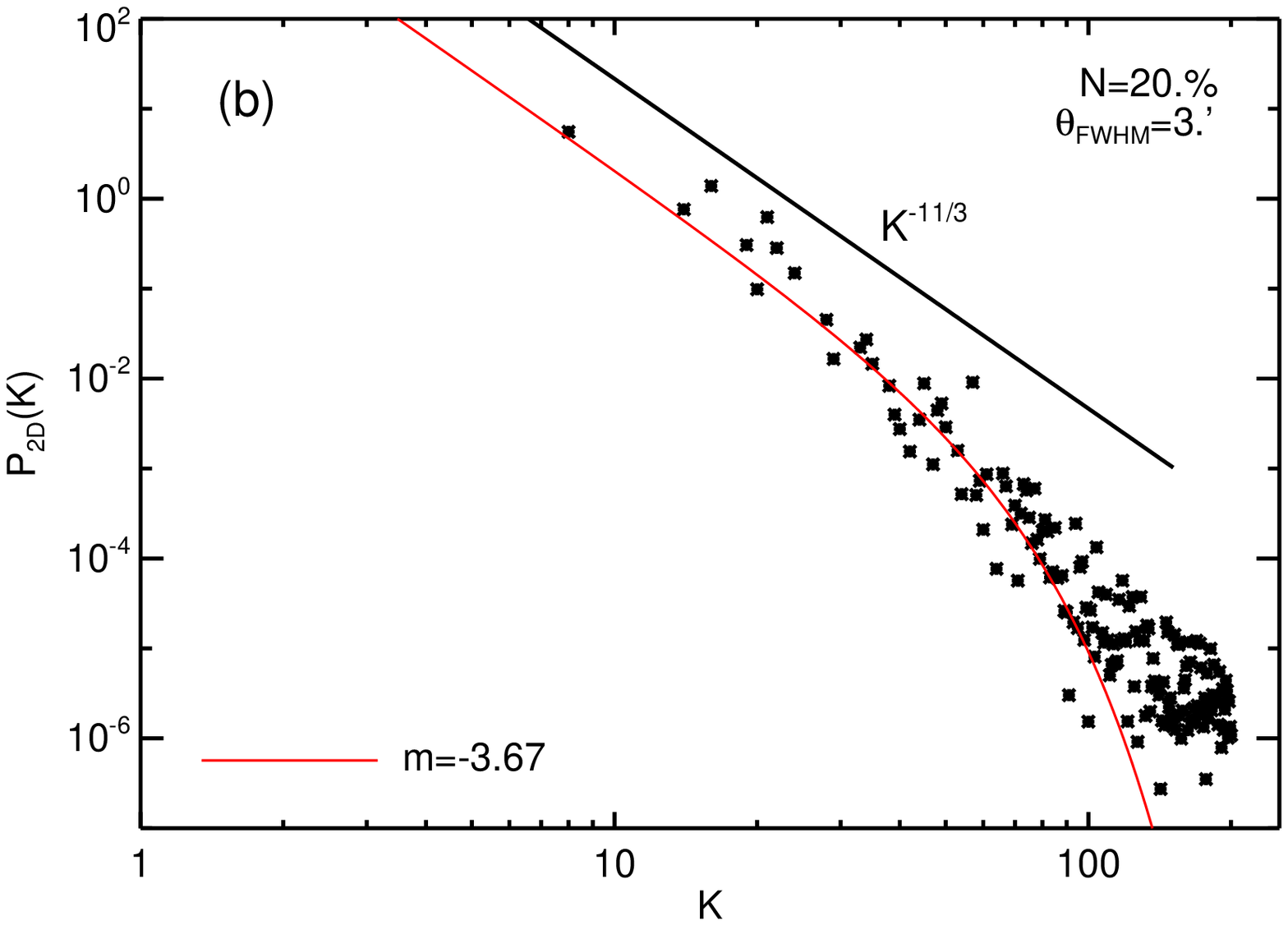}
\caption{The average 2D power spectrum from a simulated interferometric observation with noise (black $asterisks$). The level of noise is 20\% and $N_{base}=N_{base,30}$. The synthetic data are used. (a) Spectrum from an observation with a negligible beamsize. The straight red line is proportional to $K^{-11/3}$. (b) Spectrum from an observation with a finite beamsize ($\theta_{FWHM}=3'$). The red solid line is proportional to the fitting function in Equation (\ref{eq:fitting}).
}
\label{fig:synnoise}
\end{figure*}

Figure \ref{fig:synnoise} (a) shows more clearly the spectrum denoted by the red symbols (i.e.,~noise level of 20\%) in Figure \ref{fig:noise}. Due to the noise, the spectrum shows a break near $K\sim 70$. Before the break, the spectrum follows the true turbulence spectrum (see the red line) quite well.
But, after the break, the spectrum becomes flat. Note that the spectrum shown in Figure \ref{fig:synnoise} (a) is an average 2D spectrum (see Equation (\ref{eq:e2d})). Since the amplitude of noise is independent of wavenumber, the spectrum after the break is flat. As we can see in the figure, we still have approximately one decade of power-law inertial range for the noise level of $\sim 20\%$. Therefore, we can recover the true turbulence spectrum via proper fitting in the inertial range if the beam size is negligibly small and the noise level is not substantially large.

Even with a finite beam size and a noise, we may be able to recover the true turbulence spectrum. Figure \ref{fig:synnoise}(b) demonstrates this possibility. The spectrum in the figure is obtained from the following procedure. We assume that the angular size of the observed patch in the sky is $\sim 6^\circ \times 6^\circ $ and generate a synthetic polarization map at $\lambda=1cm$ on a grid of $512^2$. We also assume an underlying turbulence with a Kolmogorov spectrum (i.e.,~$k^{-11/3}$). We then smooth the map using a Gaussian beam with $\theta_{FWHM}=3^\prime$. Finally, we add a noise whose level is $20\%$ of the true polarization signal. Other observational set-up is similar to the one described earlier in this subsubsection. The resulting spectrum is shown in Figure \ref{fig:synnoise}(b). Due to a finite beam size, the spectrum does not follow a power law. Instead, it declines fast at large wavenumbers. Although the spectrum does not show a power law, we can estimate the turbulence spectrum with a fitting function of the form 
\begin{equation} \label{eq:fitting}
K^{m}e^{-K^2/2\sigma_{K}^{2}},
\end{equation} 
where $\sigma_{K}$ is the standard deviation in Fourier space defined as $1/(\sqrt{2\pi} \sigma_{beam})$. By changing the power-law index, \textit{m}, we can find the turbulence spectral index. The red solid line in Figure \ref{fig:synnoise}(a) is for $m=-11/3$ and fits the observed spectrum of the synthetic data better than other curves corresponding to different \textit{m} values.  Therefore, it is possible to recover the turbulence spectral index ($m=-11/3$) using the fitting function even in the case of finite beam size.

\subsubsection{Effects of the noise and finite beam size: simulations with MHD turbulence data} \label{sect:noise_mhd}

\begin{figure*}[h]
\centering
\includegraphics[scale=.52,angle=0]{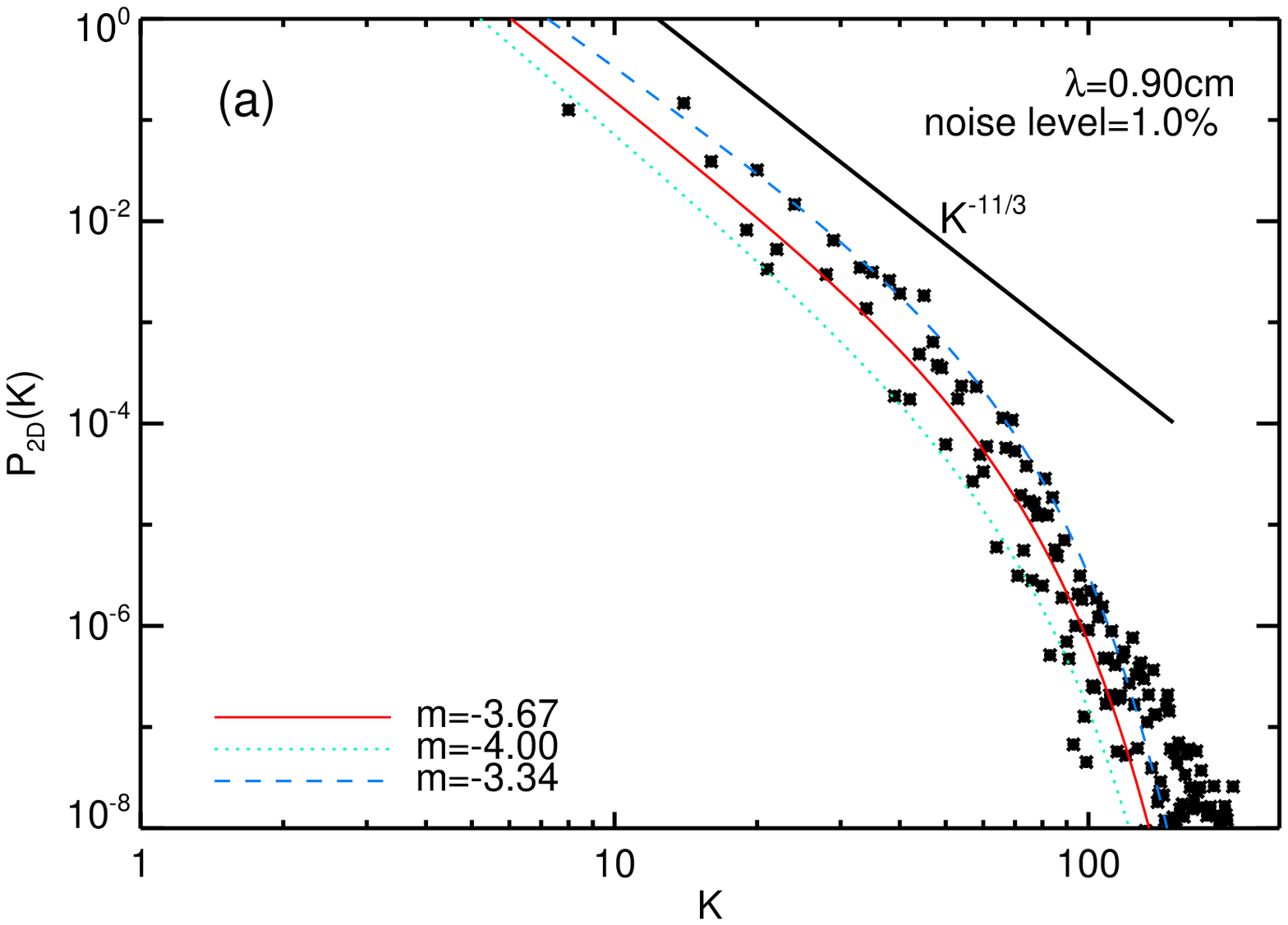}
\includegraphics[scale=.52,angle=0]{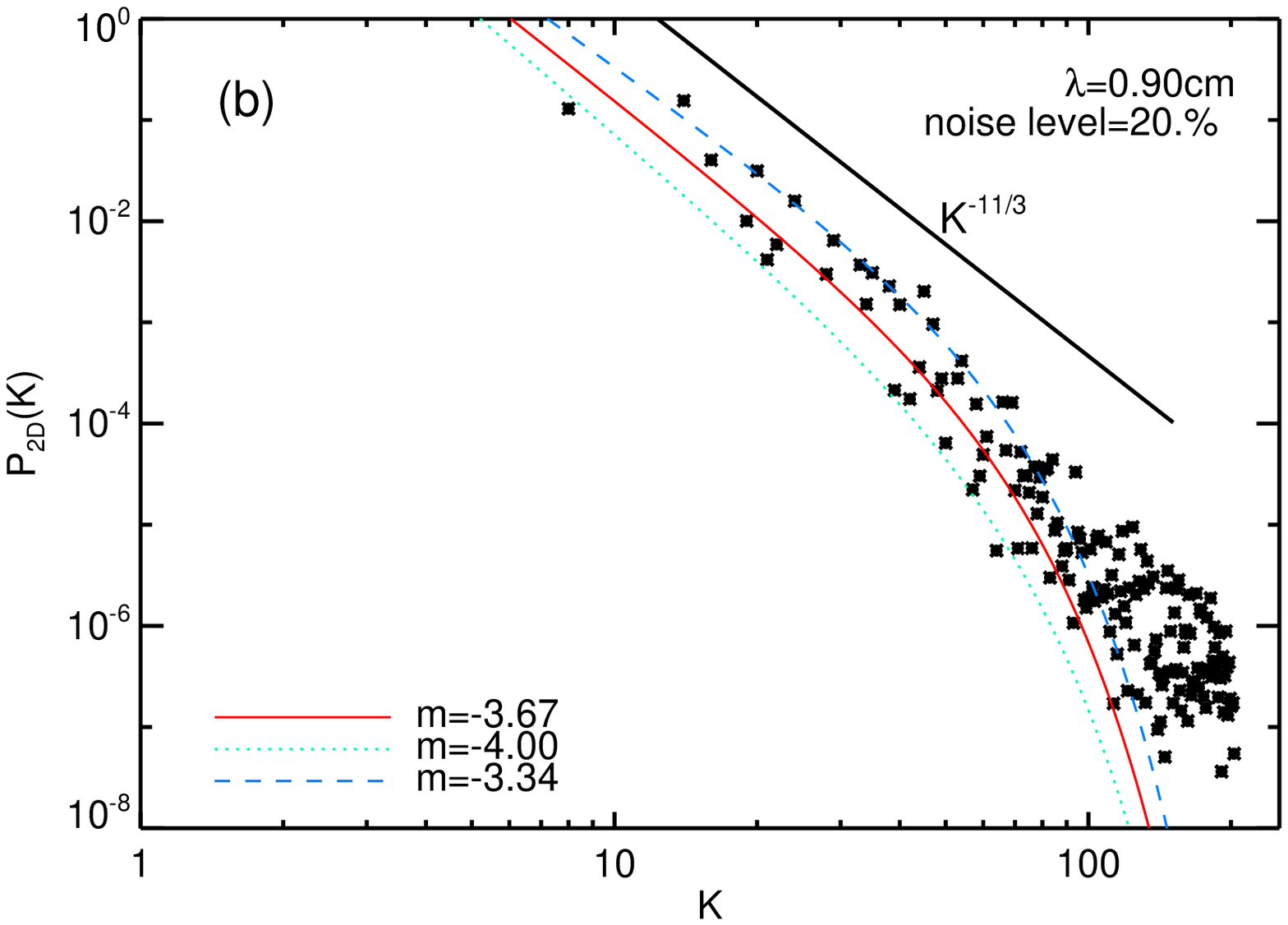}
\caption{The average 2D power spectra using trans-Alfvenic ($M_{A}\sim 0.73$) turbulence data with two different noise levels. (a) The noise level is 1\%. (b) The noise level is 20\%. Lines denote the fitting function in Equation (\ref{eq:fitting}) with different \textit{m} values: $m=-11/3$ (red solid line), $m=-12/3$ (green dotted line), and $m=-10/3$ (blue dashed line).
}
\label{fig:turbnoise}
\end{figure*}

So far we have used synthetic data for calculations. To check whether or not the fitting procedure also works for more realistic turbulence data, we use actual MHD turbulence simulation data and repeat the fitting procedure described in the previous subsubsection. The numerical method for generating the turbulence data is described in Section \ref{sect:simulation}. Both the sonic and the Alfven Mach number for the simulation are $\sim 0.7$. As in the case of synthetic data, we calculate synchrotron polarization map, add a random white noise, smooth the signal with a 3$^\prime$ Gaussian beam, and obtain the complete 2D power spectrum in wave-vector space. Then we randomly select $N_{base,30}=30 \cdot 29/2$ wave-vectors and calculate average 2D power spectrum based on the power at the selected wave-vectors (see Equation (\ref{eq:averagepower})) to mimic an interferometric observation with $N_{base,30}$ baselines. Figure \ref{fig:turbnoise} shows the resulting average 2D power spectra. Each panel in Figure \ref{fig:turbnoise} has different noise levels. The levels of noise in the left and the right panels are 1\% and 20\% of the true signal, respectively. The solid curve in the left panel represents the complete 2D power spectrum without noise and without smoothing. The spectrum shows about one decade of inertial range for $K < 40$ and the spectral index in the inertial range is compatible with $-11/3$. The filled squares in the left panel denote the simulated average 2D power spectrum based on the selected wave-vectors. Recollect that the simulated power spectrum is obtained with noise and also with smoothing. Note that the smoothing wipes out only small-scale features. In our MHD simulation, the numerical resolution is $512^3$ and we have assumed that a side of the observed sky is $\sim 6^\circ$ and the beam size is $3^\prime$. Therefore, the smoothing gets rid of features on scales smaller than a few grid zones. Since the noise level is very low and the smoothing affects spectrum only at large wavenumbers, the simulated power spectrum follows well the true complete 2D power spectrum (without noise and without smoothing) for small wavenumbers  (i.e., $ K < 80$). The filled squares in Figure \ref{fig:turbnoise}(b) denote simulated average 2D power spectrum. The red solid line in Figure \ref{fig:turbnoise}(b) is the function $K^{m}e^{-K^{2}/2\sigma_{K}^{2}}$ with  $m=-11/3$ and fits the observed 2D power spectrum fairly well. Therefore, it is possible to recover the turbulence spectral index ($m\approx -11/3$) using the fitting function even in the case of real MHD data.

\section{Discussion} \label{sect:discussion}

\subsection{Numerical tests of theory}

\subsubsection{Polarization from spatially-coincident synchrotron emission and Faraday rotation regions}

In the paper we successfully tested the predictions of LP16 that using synchrotron polarization fluctuations it is possible to recover both magnetic field and density statistics. We tested both the predictions for the spectrum of polarization $P$ and its derivative $dP/d\lambda^2$ and showed that these statistics are complimentary. In particular, the latter is focused more on the fluctuations of Faraday rotation.\footnote{The measure $dp/d\lambda^{2}$ is shown in LP16 to recover the statistics of Faraday rotation in the case of weak Faraday rotation. In this limit LP16 showed that the correlations of polarization recover only the statistics of the underlying statistics of magnetic field responsible for synchrotron fluctuations.}

For our testing we used synthetic data the spectrum of which we varied to test the theoretical predictions. We explored the effects of varying the wavelength of measurements on our studies of turbulence statistics. We found a number of numerical effects related to the finite numerical resolution of our synthetic turbulent datasets. In particular, we found that  when $\lambda^{2} \sim \frac{K_{max}}{2\pi\left<n_{e}\left| \textbf{B}_{\parallel}\right|\right>}$, structures are decorrelated with small scales and derive depolarization showing that the spectrum of polarization is getting proportional to $K$.

We tested theoretical predictions in the most complicated case considered in LP16, namely, when the volume that is emitting polarized synchrotron emission is also providing Faraday rotation due to the presence of electrons. The cases when only one effect is present in the turbulent volume are the limiting cases of the condition that we tested\footnote{The fact that these cases are simpler does not mean that they are not interesting. In a variety of astrophysically relevant setting the different regions may be dominated either by emission or Faraday rotation rather than their mixture. For instance, in a recent paper Xu \& Zhang (2016) successfully used LP16 theory to re-analyze the Faraday polarization data. They provided an interpretation of the data that is different from earlier interpretations based on the ad hoc approximations of thin Faraday screen.}.

In addition, we tested the prediction in an earlier paper by LP12 that the slope of the synchrotron spectrum does not depend on the spectral index $\gamma$ of relativistic electrons. Similar studies for synchrotron intensities were performed in Herron et al. (2016), while our study is dealing with the fluctuations of synchrotron polarization. This testing is important as $\gamma$ varies in astrophysical objects. Our results show that using expressions in LP12 one can study turbulence in synchrotron emitting volumes for arbitrary $\gamma$.

\subsubsection{Polarization from spatially-separated synchrotron emission and Faraday rotation regions}

\begin{figure}[th]
\centering
\includegraphics[scale=.52,angle=0]{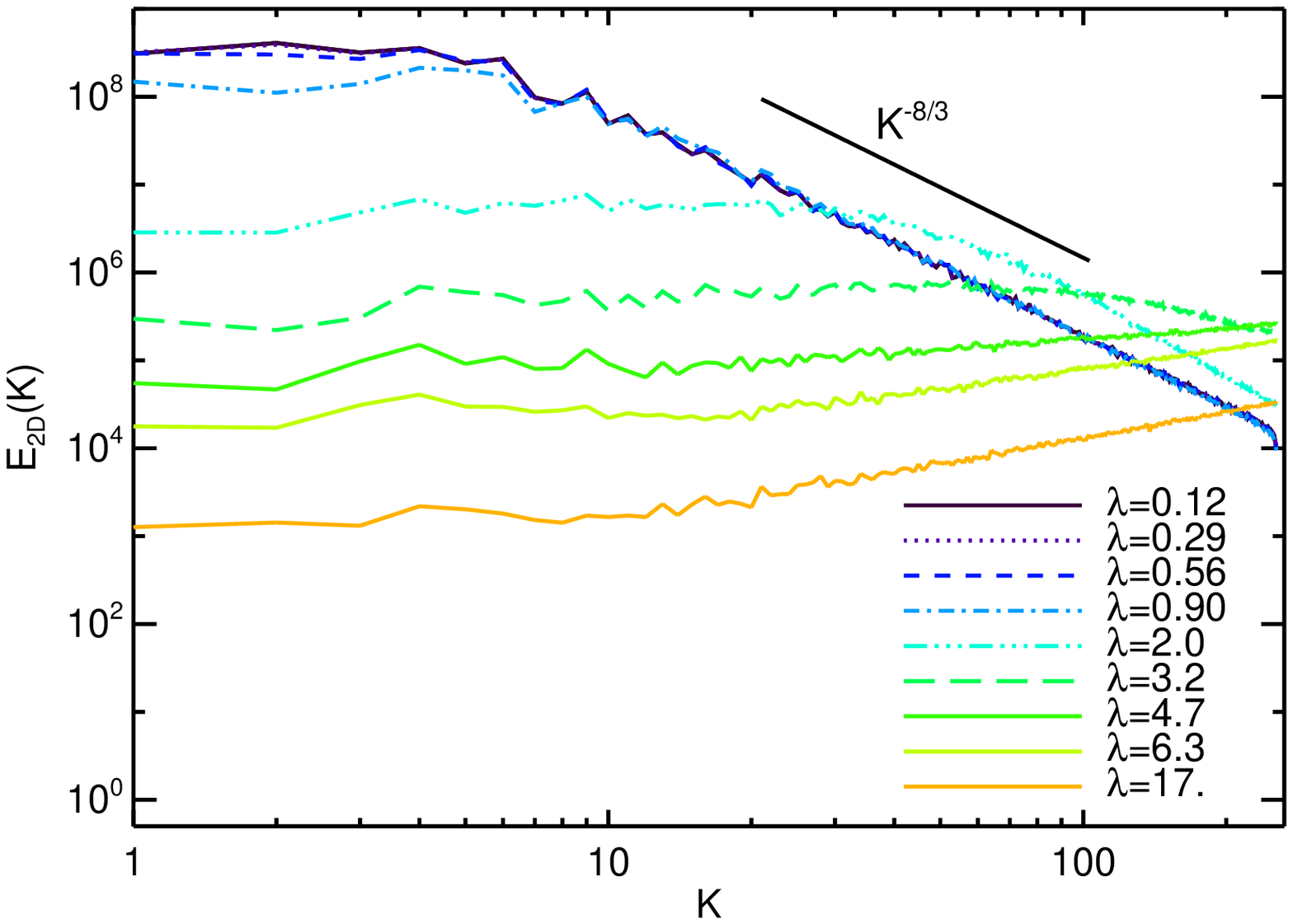}
\caption{\textit{Ring-integrated} 1D spectrum $E_{2D}(K)$ arising from separated regions of polarized synchrotron emission and Faraday rotation.  The 3D power spectrum $P_{3D}(k)$ is proportional to $k^{-11/3}$ for both electron number density and \textbf{B} in the foreground medium that produces Faraday rotation. Different curves correspond to different wavelengths. 
}
\label{fig:tworegion}
\end{figure}

In this paper, we did not consider the case in which synchrotron emitting region and Faraday rotation region are spatially separated. However, there can be a situation when synchrotron originates in one distinct region while Faraday rotation acts on synchrotron radiation in another region. Figure \ref{fig:tworegion} shows the spectra arising from separated regions of polarized synchrotron emission and Faraday rotation. We assume that background polarized synchrotron radiation is wavelength-independent and passes through a foreground medium that produces Faraday rotation. To obtain the background polarized radiation, we first generate 3D density and magnetic field on a grid of $512^{3}$ points (see section \ref{sect:simulation} for details) and calculate polarized synchrotron emission without Faraday rotation. The result of the calculation is polarized radiation on a grid of $512^{2}$, which is used as background polarized synchrotron radiation. The spectrum of the radiation follows Kolmogorov one (see the black solid curve). We also generate foreground density and magnetic field on a grid of $512^{3}$ points using different seeds for random numbers. We set $B_{0}=1$ along the LOS in the foreground medium and Faraday rotation is dominated by the mean field. The spectra of foreground magnetic field and density also follow Kolmogorov ones. We can clearly see depolarization in Figure \ref{fig:tworegion}. In fact, the overall behavior of spectra in Figure \ref{fig:tworegion} is very similar to that in Figure \ref{fig:synFR}(a). Therefore, roughly speaking, the results for spatially-coincident case can be applicable to the case in which synchrotron emitting region and Faraday rotation region are spatially separated. Note, however, that observed spectrum of polarized synchrotron emission arising from the spatially-separated regions may be more complicated. For example, the width of the Faraday rotation region may also affect the observed spectrum (see LP16).

\subsection{Importance of synchrotron studies}

Magnetic turbulence is essential for key astrophysical processes. Thus a number of techniques have been suggested to study it. While some of them are purely empirical, i.e., based on the comparison of the synthetic observations and the underlying turbulence within the numerical simulations (see  Rosolowsky et al. 1999, Padoan et al. 2001, Brunt et al. 2003, Heyer et al. 2008, Gaensler et al. 2011, Toefflemire et al. 2011, Burkhart et al. 2012, Brunt \& Heyer 2013, Burkhart \& Lazarian 2015),  others are based on the theoretical description of turbulence statistics. For instance, in a series of papers by Lazarian \& Pogosyan (2000, 2004, 2006, 2008) the statistics of intensity fluctuations within Position-Position-Velocity (PPV) data cutes was described for the Doppler shifted spectral lines from turbulent volumes. These studies provide the way to recover the statistics of density and velocity (see Lazarian 2009 for a review as well as Padoan et al. 2009, Chepurnov et al. 2010, 2015). The magnetic field statistics provides the complimentary essential piece of information and the studies in LP16 were aimed at obtaining a theory-motivated way to study magnetic turbulence from observations. Our successful testing of some of the suggested techniques paves the way to the application of these techniques to observational data.

Our study is complementary to that in a recent paper by Zhang et al. (2016), where the other analytical expressions from LP16 were tested. Indeed, Zhang et al. (2016) tested the fluctuations of the polarization variance with the wavelength. The input data for such studies is the polarized synchrotron radiation collected along a single line of sight but with the wavelength being changed. In contrast, in this paper we studied the statistics of two point correlations for the same wavelength.

Studying synchrotron variations is not only important for astrophysical applications, e.g., for understanding better processes of cosmic ray propagation, transport of heat, star formation etc., but also for observational cosmology. Indeed, synchrotron polarization fluctuation presents an important foreground for search of enigmatic B-modes of cosmological origin. Thus our testing of how these fluctuation varies with the wavelength as well as with the variations of $\gamma$ are of prime applicability to such a search.

\subsection{Complementary ways of studying turbulence}

Our present work opens ways for studying turbulence using synchrotron polarization. Such studies are complementary
to the spectroscopic Doppler-shift studies of fluctuations. Those studies can be performed using Doppler shifted  lines
using Velocity Channel Analysis (VCA, Lazarian \& Pogosyan 2000, 2004; Kandel, Lazarian \& Pogosyan 2016a), Velocity Coordinate Spectrum (VCS, Lazarian \& Pogosyan 2006, 2008) and Velocity Centroids (Lazarian \& Esquivel 2003; Esquivel \& Lazarian 2005, 2010,
Burkhart et al. 2014; Kandel, Lazarian \& Pogosyan 2016; Cho \& Yoo 2016).

Combining the techniques it is possible to see how the turbulence varies from one media, that can be sampled by spectral lines to another, that can be sampled by synchrotron fluctuations. This can answer important questions related to whether the turbulence in the interstellar medium presents on big cascade or whether different phases of the ISM maintain their individual turbulent cascades.

\section{Summary} \label{sect:summary}

We successfully tested analytical expressions in LP16 and proved that the techniques suggested there 
can be used to analyze observed fluctuation of polarized synchrotron radiation in order to restore the statistics of underlying magnetic turbulence. Our numerical results testify that such a study can be performed 
\begin{itemize}
\item for arbitrary spectral index of relativistic electrons,
\item in the presence of Faraday rotation and depolarization  caused by turbulent magnetic field,
\item in the settings  when only Faraday rotation is responsible for the polarization fluctuations,
\item in the presence of effects of finite beamsize and noise, and
\item in case the data are obtained with an interferometer with measurement performed for just a few baselines.
\end{itemize}
We believe that our present study paves the way for the successful use of LP16 techniques with observational data.
\acknowledgements
HL thanks the Dept. of Astronomy, University of Wisconsin-Madison for the hospitality.  A. L. acknowledges the NSF grant AST 1212096 and Center for Magnetic Self Organization (CMSO). HL and JC's work is supported by the National R \& D Program through the National Research Foundation of Korea, funded by the Ministry of Education (NRF-2013R1A1A2064475).


\clearpage
\appendix
\section{A. Synchrotron emission and the stokes parameters}
The study of intensity and linear polarization of the synchrotron emission provides a valuable information on the magnetic field. The observed intensity and polarized emission can be described by:
\begin{equation}
I=\int d\Omega\, \text{I},\quad P\equiv Q+iU=\int d\Omega\,\text{P}(\textbf{X},\lambda^{2}) 
\end{equation}
where \textit{I} is the specific intensity and \text{P}(\textbf{X},$\lambda^{2}$) is observed polarized intensity given by

\begin{eqnarray} 
\text{I}&=&\int_{0}^{L}\,\left(j_{\perp}(\omega,\textbf{x})+j_{\parallel}(\omega,\textbf{x})\right)dz \nonumber \\
\text{P}\left(\textbf{X},\lambda ^{2}\right)&=&\int_{0}^{L}\,\left(j_{\perp}(\omega,\textbf{x})-j_{\parallel}(\omega,\textbf{x})\right) e^{2i\chi(\textbf{X},z)} dz,
\end{eqnarray}
where $\textbf{x}=(\textbf{X},z)$, $j_{\perp}$ and $j_{\parallel}$ are the synchrotron emissivity perpendicular and parallel to $B_{\perp}$, respectively (Waelkens et al 2009):
\begin{eqnarray} 
j_{\perp}(\omega,\textbf{x}) = [F(p)+G(p)]\omega^{\frac{1-p}{2}}\left| \textbf{B}_{\perp}(\textbf{x}) \right| \nonumber \\
j_{\parallel}(\omega,\textbf{x}) = [F(p)-G(p)]\omega^{\frac{1-p}{2}}\left| \textbf{B}_{\perp}(\textbf{x}) \right|
\end{eqnarray}

The polarization angle with respect to the plane of the sky is
\begin{equation} \label{eq:angle}
\chi(\textbf{X},z)=\chi_{0}+\lambda^{2}\Phi(\textbf{X},z)
\end{equation}
where $\chi_{0}$ is the intrinsic polarization angle:
\begin{equation}
\chi_{0}=\text{tan}^{-1}\left(\frac{B_{y}}{B_{x}}\right)
\end{equation}
and $\Phi(\textbf{X},z)$ is the Faraday rotation measure (see Equation \ref{eq:RM}).

The polarization can be represented by the Stokes vector, $\textbf{S}=\left(I,Q,U\right)$, where I is the total intensity, and Q and U describe the linear polarization. The Stokes parameters neglecting Faraday rotation $(\Phi(\textbf{X},z)=0)$ are defined as follows:
\begin{eqnarray}
I&=&\enspace\,\, 2F(p)\omega^{\frac{1-p}{2}}\int d\Omega\int dz{\left({B_{x}}^{2}+{B_{y}}^{2}\right)}^{\frac{p-3}{4}}\left({B_{x}}^{2}+{B_{y}}^{2}\right)\nonumber\\
Q&=&-2G(p)\omega^{\frac{1-p}{2}}\int d\Omega\int dz{\left({B_{x}}^{2}+{B_{y}}^{2}\right)}^{\frac{p-3}{4}}\left({B_{x}}^{2}-{B_{y}}^{2}\right)\\
U&=&-2G(p)\omega^{\frac{1-p}{2}}\int d\Omega\int dz{\left({B_{x}}^{2}+{B_{y}}^{2}\right)}^{\frac{p-3}{4}}2\left(B_{x}B_{y}\right)\nonumber
\end{eqnarray}
where $\omega=2\pi c/\lambda$, $\lambda$ is the observation wavelength, p is the spectral index of the electron energy  distribution, and

\begin{equation}
\begin{array} {l}
\displaystyle F\left(p\right)=\frac{\sqrt{3\pi}e^{3}N_{0}}{64\pi^{2}c^{2}m_{e}}\left(\frac{2m_{e}c}{3e}\right)^{\frac{1-p}{2}}\frac{2^{\frac{p+1}{2}}}{p+1}\frac{\Gamma\left(\frac{p}{4}+\frac{19}{12}\right)\Gamma\left(\frac{p}{4}-\frac{1}{12}\right)\Gamma\left(\frac{p}{4}+\frac{5}{4}\right)}{\Gamma\left(\frac{p}{4}+\frac{7}{4}\right)}\\
\displaystyle G\left(p\right)=\frac{\sqrt{3\pi}e^{3}N_{0}}{64\pi^{2}c^{2}m_{e}}\left(\frac{2m_{e}c}{3e}\right)^{\frac{1-p}{2}}2^{\frac{p-3}{2}}\frac{\Gamma\left(\frac{p}{4}+\frac{7}{12}\right)\Gamma\left(\frac{p}{4}-\frac{1}{12}\right)\Gamma\left(\frac{p}{4}+\frac{5}{4}\right)}{\Gamma\left(\frac{p}{4}+\frac{7}{4}\right)}.
\end{array}
\end{equation}
Here $m_{e}$ is the electron mass, e is its charge, and $N_{0}$ is the pre-factor of the electron distribution. 

\section{B. Comparison of Two Point Statistics : power spectra and structure function}
Since the power spectrum can provide information on energy transfer processes in turbulence, accurate measurement of the power spectrum is essential for our understanding of astrophysical turbulence. The power spectrum can be obtained from the Fourier transform of the correlation function, $CF(\textbf{r})\equiv \left<P(\textbf{x})P(\textbf{x}+\textbf{r}) \right>$:
\begin{equation}
P_{3D}(\textbf{k}) = \frac{1}{(2\pi)^{3}} \int {\left<P(\textbf{x})P(\textbf{x}+\textbf{r}) \right> e^{-i\textbf{k}\cdot\textbf{r}}d\textbf{r}},
\end{equation}
where $<...>$ denotes average over \textbf{x}.
Here $P_{3D}(\textbf{k})$ is the 3D power spectrum.
In this paper, we use the following definitions for different types of spectrum.
\begin{enumerate}
\item $P_{3D}(\textbf{k})$: the 3D power spectrum. 
              $P_{3D}(\textbf{k})=| \tilde{ \textbf{v} }_{\textbf{k}} |^2$, where $\tilde{ \textbf{v} }_{\textbf{k}}$ is the 3D Fourier transform
of a real space variable $\textbf{v}(\textbf{x})$. 

\item In case of a 2D observable, we also use $P_{2D}(\textbf{K})$ for the 2D power spectrum. In this case,
              $P_{2D}(\textbf{K})=| \tilde{ \textbf{S} }_{\textbf{K}} |^2$, where $\tilde{ \textbf{S} }_{\textbf{K}}$ is the 2D Fourier transform
of a 2D real space variable $\textbf{S}(\textbf{x})$. If $S=\int s dl$, where $s$ is a variable in 3D space and the integration is done along the LOS, the 2D power spectrum of $S$ is proportional to the 3D power spectrum of $s$ if the domain is periodic in the direction of the LOS.

\item $E_{3D}(k)$: the \textit{shell-integrated} 1D spectrum for a 3D variable. $E_{3D}(k)\equiv \int_{k-0.5}^{k+0.5} P_{3D}(k) dk$, where
        the integration is done over a shell of radius $k$ in 3D Fourier space and $P_{3D}(k)$ is the 3D power spectrum. 
        If turbulence is isotropic and $P_{3D}(k)\propto k^{m}$, then the shell-integrated 1D spectrum $E_{3D}(k)$ is proportional to $k^{m+2}$.
        
\item $E_{2D}(K)$: the \textit{ring-integrated} 1D spectrum for a 2D variable. $E_{2D}(K)\equiv \int_{K-0.5}^{K+0.5} P_{2D}(K) dK$, where
        the integration is done over a ring of radius $K$ in 2D Fourier space and $P_{2D}(K)$ is the 2D power spectrum.
        If the 2D spectrum is isotropic and $P_{2D}(K)\propto K^{m}$, then the next $E_{2D}(K)$ is proportional to $K^{m+1}$.

\end{enumerate}

In some cases, it is easier to obtain the second-order structure function from observations than the spectrum. The power-law indices for the former and the latter are related as follows.
For both 2D and 3D cases, if the spectral index of the 1D spectrum is $\alpha$ (i.e.,~$E(k)\propto k^{\alpha}$), then
the power-law index for the second-order structure function is $-\alpha-1$ (i.e.,~$SF_2(r) \propto r^{-\alpha-1}$):
\begin{equation}
   E(k)\propto k^{-\alpha}  ~~ \rightarrow ~~SF_2(r) \propto r^{\alpha-1}  ~~ (\alpha < 3)
\end{equation}
(see, for example, Cho \& Lazarian 2009). Here, the 1D spectrum can be either shell-integrated 1D spectrum for 3D data or the ring-integrated 1D spectrum for 2D data.
Therefore, we can indirectly obtain the spectral index of spectrum using the second-order structure function. 
The second-order structure function in either 2D or 3D can be directly obtained by
\begin{equation}
               \text{SF}_{2}(\textbf{r}) = \left\langle [P(\textbf{x}) - P(\textbf{x}+\textbf{r})]^{2} \right\rangle,
\end{equation}
where $\textbf{r}$ is the separation vector in either 2D or 3D. 
Or, given a spectrum (or a correlation function $CF(\textbf{r})=\left<P\left(\textbf{x})P(\textbf{x}+\textbf{r}\right)\right>$), the second-order structure function can be calculated by
\begin{equation}
\text{SF}_{2}(\textbf{r}) = 2[\text{CF}(0) - \text{CF}(\textbf{r})].   \label{eq:sf2cf}
\end{equation}

\section{C. Limitation of the second-order structure function}

\begin{figure*}[h]
\centering
\includegraphics[scale=.52,angle=0]{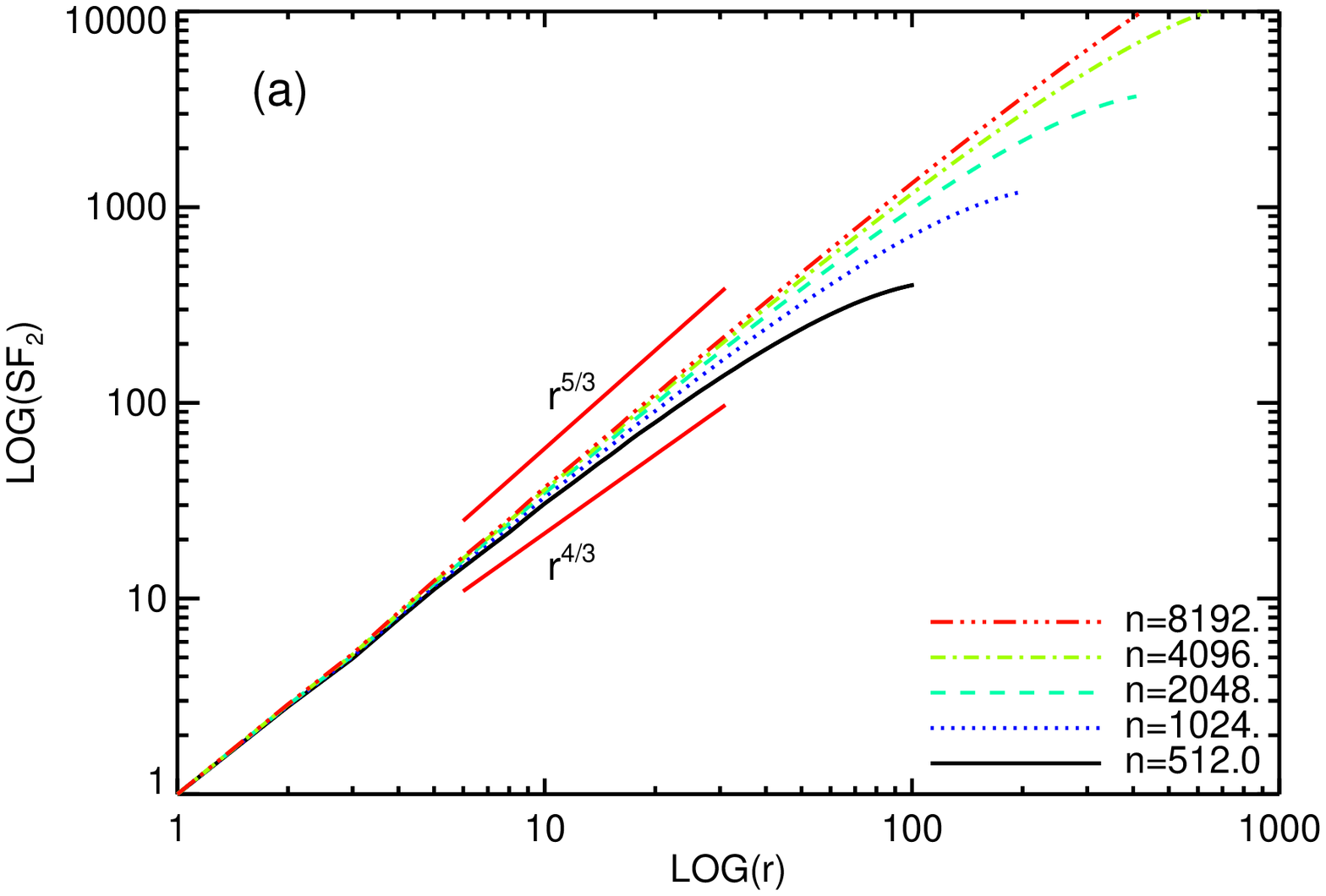}
\includegraphics[scale=.52,angle=0]{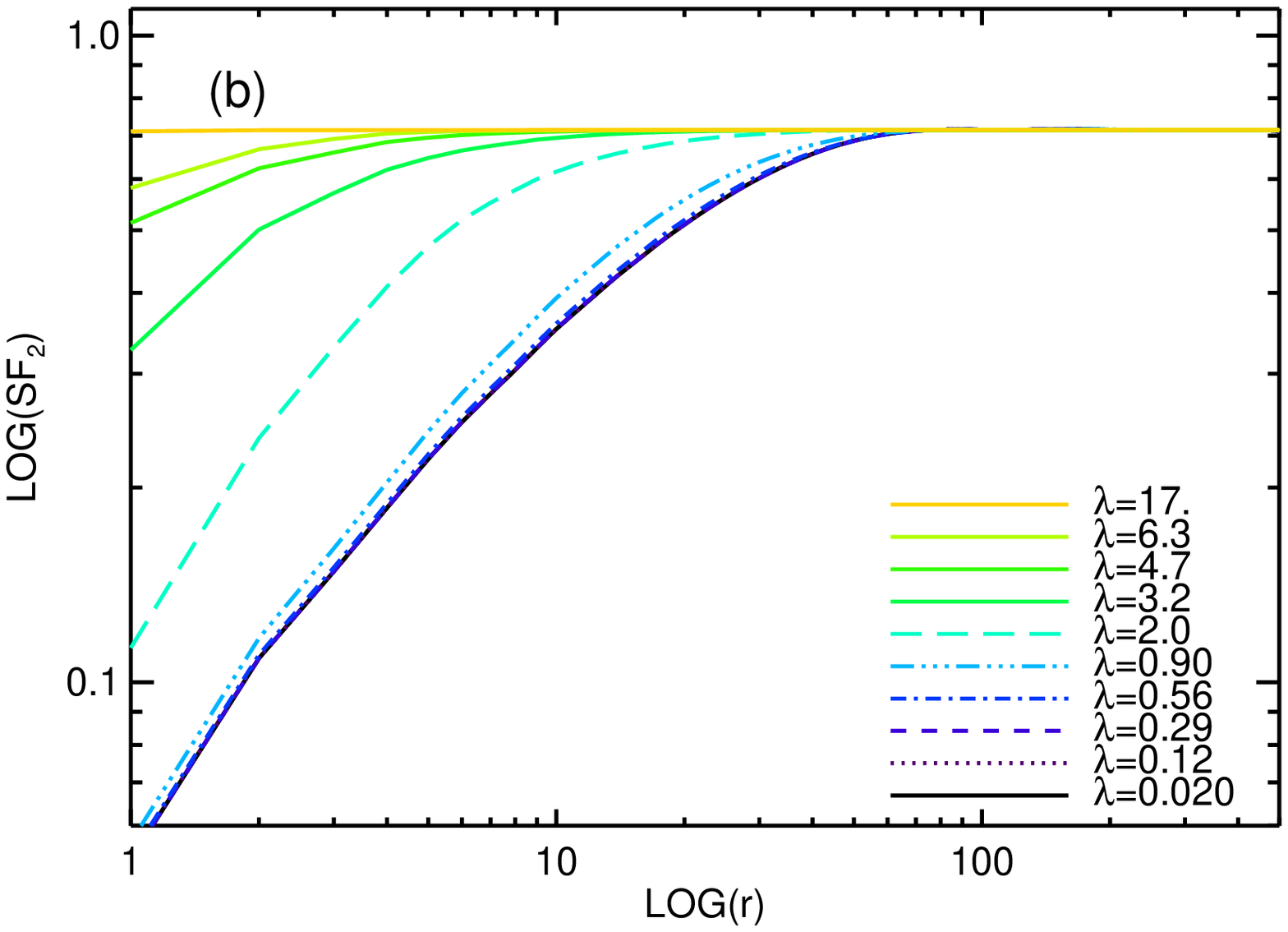}
\caption{(a) The second-order structure functions calculated directly from 2D synthetic data with resolution of $512^2$, $1024^2$, $2048^2$, $4096^2$, and $8192^2$. (b) The second-order structure function of polarized synchrotron emission from the synthetic data. The structure functions correspond to spectra in Figure \ref{fig:synFR}(a). Different curves denote different wavelengths. 
}
\label{fig:sf2d}
\end{figure*}

Note that the use of the second-order structure function to obtain the true turbulence spectrum requires a very high numerical resolution.
In this Appendix, we demonstrate this limitation of the structure function.

We first generate an isotropic 2D power spectrum for a scalar variable in Fourier space, which follows a $K^{-11/3}$ spectrum between
$K=2$ and $K=K_{max}-2$, where $K_{max}$ is equal to the half of the numerical resolution $N$.
We use $N=$ 512, 1024, 2048, 4096, and 8192.
Then, we calculate the second-order structure function using Equation (\ref{eq:sf2cf}).
Since the 2D power spectrum is isotropic, the second-order structure function depends on the scalar separation $r$.

We plot the resulting second-order structure function in Figure \ref{fig:sf2d}(a).
Since the \textit{ring-integrated} 1D spectrum $E_{2D}(K)$ is proportional to $K^{-8/3}$, we expect that $SF_2(r) \propto r^{5/3}$.
However, according to the figure, the second-order structure function does not shows the expected scaling relation.
When $N=512$, the power-law slope of the second-order structure function lies between $4/3$ and $5/3$.
As numerical resolution $N$ increases, the slope becomes steeper.
But, even with $8192^2$ resolution, the slope is still less than $5/3$.
Therefore, it is not easy to reveal the true turbulence spectrum using the second-order structure function unless the resolution is very high.

Indeed, if we plot the second-order structure function of synchrotron polarization for the case that both density and magnetic field 
have Kolmogorov spectra (see Figure \ref{fig:synFR}(a)), we get slope less than the expected one.
In Figure \ref{fig:sf2d}(b), we plot the corresponding second-order structure function for different wavelength, $\lambda$. We expect that the second-order structure functions are proportional to $r^{5/3}$ for small spatial separation, since the spectral index of the \textit{ring-integrated} 1D spectrum $E_{2D}(K)$ is $-8/3$ at large $K$. However, Figure \ref{fig:sf2d}(b) shows that the slopes of second-order functions do not follow the expectation. 


\begin{thebibliography}{}

\bibitem[a(2016)]{1995ApJ443,209} Armstrong, J. M., Rickett, B. J., the, S. R., 1995, ApJ, 443, 209
\bibitem[a(2016)]{2002arXiv,0203353v1} Balbus, S. A., \& Hawley, J. F., 2002, arXiv0203353v1
\bibitem[a(2016)]{2015AA,578,A93} Beck, R. 2015, A\&A, 578, A93
\bibitem[a(2016)]{2013arXiv,1302.5663} Beck, R. Wielebinski, R. 2013, arXiv1302.5663
\bibitem[a(2016)]{2015} Beresnyak A., \& Lazarian A., 2015, in , Magnetic Fields in Diffuse Media. Springer, pp 163?226
\bibitem[a(2016)]{2013SSRv,178,163} Brandenburg, A., \& Lazarian, A., 2013, SSRv, 178, 163
\bibitem[a(2016)]{2013MNRAS,433,117} Brunt, C. M., \& Heyer, M.H., 2013, MNRAS, 433, 117
\bibitem[a(2016)]{2003ApJ,595,824} Brunt, C.M., Heyer, M.H., Vázquez-Semadeni, E., \& Pichardo, B. 2003, ApJ, 595, 824
\bibitem[a(2016)]{2015arXiv,1511.03660} Burkhart B., \& Lazarian A. 2015, arXiv.1511.03660v1
\bibitem[a(2016)]{2012ApJ,749,145} Burkhart B., Lazarian A., \& Gaensler B., 2012, ApJ, 749, 145
\bibitem[a(2016)]{2014ApJ790130B} Burkhart, B., Lazarian, A., Le{\~a}o, I.~C., de Medeiros, J.~R., \& Esquivel, A., 2014, ApJ, 790, 130 
\bibitem[a(2016)]{2015ApJ,810,33} Chepurnov A., Burkhart B., Lazarian A., Stanimirovic, S.2015,ApJ,810,33
\bibitem[a(2016)]{2009ApJ,693,1074} Chepurnov A., \& Lazarian A., 2009, ApJ, 693, 1074
\bibitem[a(2016)]{2010ApJ,714,1398} Chepurnov A., Lazarian A., Stanimirovic, S., Heiles C., Peek J. 2010, ApJ, 714, 1398
\bibitem[a(2016)]{2002ApJ575,L63} Cho, J., \& Lazarian, A. 2002a, ApJ, 575, L63
\bibitem[a(2016)]{2002Phy.Rev.Lett88, 245001} Cho, J., \& Lazarian, A. 2002b, Phy.Rev.Lett., 88, 245001
\bibitem[a(2016)]{2003,MNRAS,345,325} Cho J., \& Lazarian A., 2003, MNRAS, 345, 325
\bibitem[a(2016)]{2010ApJ,720,1181} Cho, J., \& Lazarian, A. 2010 ApJ, 720, 1181
\bibitem[a(2016)]{2002ApJ,564,291} Cho J., Lazarian A., \& Vishniac E. T., 2002, ApJ, 564, 291
\bibitem[a(2016)]{2000ApJ,539,273} Cho J., \& Vishniac E. T., 2000, ApJ, 539, 273
\bibitem[a(2016)]{2016ApJ,821,21} Cho, J., \& Yoo, H., 2016 ApJ, 821, 21
\bibitem[a(2016)]{2003PRD,68,083003} de Oliveira-Costa, A., Tegmark, M., Odell, C., Keating, B., Timbie, P., Efstathiou, G., \& Smoot, G. 2003, Phys. Rev. D, 68, 083003
\bibitem[a(2016)]{2002ApJ,577,206} Elmegreen, B. 2002, ApJ, 577, 206
\bibitem[a(2016)]{1996ApJ,471,816} Elmegreen, B., \& Falgarone, E. 1996, ApJ, 471, 816
\bibitem[a(2016)]{2004ARAA,42,211} Elmegreen, B.G., \& Scalo, J. 2004, ARA\&A, 42, 211
\bibitem[a(2016)]{2011ApJ740117E} Esquivel, A., \& Lazarian, A., 2011, ApJ, 740, 117 
\bibitem[a(2016)]{2011Nature,478,214} Gaensler, B. M., Haverkorn, M., Burkhart, B., Newton-McGee, K. J., Ekers, R. D., Lazarian, A., McClure-Griffiths, N. M., Robishaw, T., Dickey, J. M., \& Green, A. J. 2011, Nature, 478, 214
\bibitem[a(2016)]{2010BAAS,215,470.13} Gaensler, B. M., Landecker, T. L., \& Taylor, A. R. POSSUM Collaboration 2010 BAAS, 215, 470.13
\bibitem[a(2016)]{1995ApJ,438,763} Goldreich, P., \& Sridhar, S. 1995, ApJ, 438, 763
\bibitem[a(2016)]{2016arXiv,1604.02751v1} Herron, C.A.,  Burkhart B, Lazarian, A., Gaensler, B.M., \& McClure-Griffiths, N. M., 2016, arXiv1603.02751v1
\bibitem[a(2016)]{2008ApJ,680,420} Heyer, M., Gong, H., Ostriker, E., \& Brunt, C. 2008, ApJ, 680, 420
\bibitem[a(2016)]{1984ApJ,285,109} Higdon J., 1984, ApJ, 285, 109
\bibitem[a(2016)]{1964SA,7,566} Iroshnikov, P. 1964, Soviet Astronomy, 7, 566
\bibitem[a(2016)]{2016MNRAS.tmp..947K} Kandel, D., Lazarian, A., \& Pogosyan, D.\ 2016a, \mnras (in press)
\bibitem[a(2016)]{2016MNRAS.tmp..947K} Kandel, D., Lazarian, A., \& Pogosyan, D.\ 2016b, \mnras (submitted)
\bibitem[a(2016)]{1941SSSR,31,538} Kolmogorov, A. 1941, Dokl. Akad, Nauk SSSR, 31, 538
\bibitem[a(2016)]{1962}  Kolmogorov, A.N. 1962  J. Fluid Mech. 13, pp. 82-85.
\bibitem[a(2016)]{2007ApJ,666,L69} Kowal, G., \& Lazarian, A. 2007, ApJ, 666, L69
\bibitem[a(2016)]{1983MNRAS,194,809} Larson, R. B 1981, MNRAS, 194, 809
\bibitem[a(2016)] {2009SSR,143,357} Lazarian, A. 2009, Space Sci. Rev., 143, 357
\bibitem[a(2016)]{2015} Lazarian A., Eyink G., Vishniac E., Kowal G., 2015, Philosophical Transactions of the Royal Society of London A: Mathematical, Physical and Engineering Sciences, 373, 20140144
\bibitem[a(2016)]{2000ApJ,537,720} Lazarian, A., \& Pogosyan, D. 2000, ApJ, 537, 720
\bibitem[a(2016)]{2002ASPC,276,182L} Lazarian, A., Pogosyan, D., \& Esquivel, A. 2002, in Seeing Through the Dust, ed. R. Taylor, T. Landecker, \& A. Willis, ASP Conf. Series (ASPC), 276, 182L
\bibitem[a(2016)]{2004ApJ,616,934} Lazarian, A., \& Pogosyan, D. 2004, ApJ, 616, 934
\bibitem[a(2016)]{2006ApJ,652,1348} Lazarian, A., \& Pogosyan, D. 2006, ApJ, 652,1348
\bibitem[a(2016)]{2008ApJ,686,350} Lazarian, A., \& Pogosyan, D. 2008, ApJ, 686,350
\bibitem[a(2016)]{2012ApJ,747,5} Lazarian, A., \& Pogosyan, D. 2012, ApJ, 747, 5 [LP12]
\bibitem[a(2016)]{2016ApJ,818,178} Lazarian A., Pogosyan D., 2016, ApJ, 818, 178 [LP16]
\bibitem[a(2016)]{1999ApJ,517,700} Lazarian A., \& Vishniac E. T., 1999, ApJ, 517, 700
\bibitem[a(2016)]{1998JGR,103, 4775L} Leamon, R. J., Smith, C. W., Ness, N. F.,  Matthaeus, W. H., \& Wong, H. K. 1998, J. Geophys. Res., 103, 4775
\bibitem[a(2016)]{ApJ,562,279} Lithwick, Y., \& Goldreich, P. 2001, ApJ, 562, 279
\bibitem[a(2016)]{2004RVMP76,125} Mac Low, M., \& Klessen, R. 2004, RVMP, 76, 125
\bibitem[a(2016)]{2001ApJ,554,1175} Maron J., \& Goldreich P., 2001, ApJ, 554, 1175
\bibitem[a(2016)]{2007ARAA,45,565} McKee, C., \& Ostriker, E. 2007, ARA\&A, 45, 565
\bibitem[a(2016)]{2002Nature,416,59} McKee, C., \& Tan, J. 2002, Nature, 416, 59
\bibitem[a(2016)]{1981PF,24,825} Montgomery D., Turner L., 1981, Physics of Fluids, 24, 825
\bibitem[a(2016)]{1958RMP,30,1035} Munch G., 1958, Reviews of Modern Physics, 30, 1035
\bibitem[a(2016)]{1958PoF,1,462} Munch G., Wheelon A. D., 1958, Physics of Fluids (1958-1988), 1, 462
\bibitem[a(2016)]{2009ApJ,707,153} Padoan, P., Juvela, M., Kritsuk, A., \& Norman, M.L., ApJ, 707,153
\bibitem[a(2016)]{2001ApJ,547,862} Padoan P., Rosolowsky E. W., Goodman A. A., 2001, ApJ, 547, 862
\bibitem[a(2016)]{1999ApJ,524,887} Rosolowsky, E. W., Goodman, A. A., Wilner, D. J., \& Williams, J. P. 1999, ApJ, 524, 887
\bibitem[a(2016)]{2002Springer} Schlickeiser, R. 2002, Cosmic Ray Astrophysics (Berlin: Springer)
\bibitem[a(2016)]{1983JPP,29,525} Shebalin J. V., Matthaeus W. H., Montgomery D., 1983, Journal of Plasma Physics, 29, 525
\bibitem[a(2016)]{2015ApJ,149,60} Sun, X. H., Rudnick, L., Akahori, T., et al. 2015, ApJ, 149, 60
\bibitem[a(2016)]{2011,ApJ,736,60}	Tofflemire, B.M., Burkhart, B., \& Lazarian, A. 2011, ApJ, 736, 60
\bibitem[a(2016)]{2009MNRAS, 398, 1970} Waelkens, A. H., Schekochihin, A. A., \& Enblin, T. A. 2009, MNRAS, 398, 1970
\bibitem[a(2016)]{2012ApJ,746,103} Wicks, R. T., Forman, M. A., Horbury, T. S., \& Oughton S. 2012, ApJ, 746, 103
\bibitem[a(2016)]{2016arXiv} Zhang, J.F., Lazarian, A., Lee, H., Cho, J. 2016, arXiv1605.01307

\end{thebibliography}
\end{document}